\documentclass[aps, amsmath,amssymb,twocolumn,pra]{revtex4}
\usepackage{graphicx}
\usepackage{dcolumn}
\usepackage{bm}
\usepackage{hyperref}
\usepackage{color}  

\begin{document}

\title{Dissipative Effects on the Superfluid to Insulator Transition \\
in Mixed-dimensional Optical Lattices}
\author{E. Malatsetxebarria}
\affiliation{
Centro de F\'isica de Materiales (CFM), Centro Mixto CSIC-EHU, Paseo Manuel de Lardizabal 5
E-20018 San Sebastian, Spain.}
\affiliation{Donostia International Physics Center (DIPC), Paseo Manuel de Lardizabal 4, E-20018 San Sebastian, Spain.}
\author{Zi Cai}
\author{U. Schollw\"{o}ck}
\affiliation{Department of Physics and Arnold Sommerfeld Center for
Theoretical Physics, Ludwig-Maximilians-Universit\"{a}t M\"{u}nchen,
D-80333 M\"{u}nchen, Germany}
\author{ Miguel A. Cazalilla}
\address{Graphene Research Centre National University of Singapore, 6 Science
Drive 2, Singapore 117546}
\affiliation{
Centro de F\'isica de Materiales (CFM), Centro Mixto CSIC-EHU, Paseo Manuel de Lardizabal
E-20018 San Sebastian, Spain.}
\affiliation{Donostia International Physics Center (DIPC), Paseo Manuel de Lardizabal 4, E-20018 San Sebastian, Spain.}
\begin{abstract}
We study the superfluid to Mott insulator transition of a mixture of heavy bosons and light fermions loaded in an optical
lattice.    We focus on the effect of the light fermions on the dynamics of the heavy bosons. It is shown that, when the lattice
potential is sufficiently deep to confine the bosons to one dimension but allowing the fermions  to freely move in three
dimensions (i.e. a mixed-dimensionality lattice), the fermions  act as an ohmic bath 
for bosons  leading to screening  and dissipation effects on the
bosons. Using a perturbative renormalization-group analysis, it is shown that the fermion-induced 
dissipative effects have no appreciable impact on the transition from  the superfluid to the Mott-insulator state at integer filling.
On the other hand, dissipative effects are found to be very  important in the half-filled case near the critical point. In this case,  in the presence of a finite incommensurability that destabilizes the Mott phase, the bosons can still be localized by virtue of dissipative 
effects.
\end{abstract}

 \date{\today}
\maketitle

\section{Introduction}

\begin{figure}[b]
\includegraphics[height=0.30\textwidth]{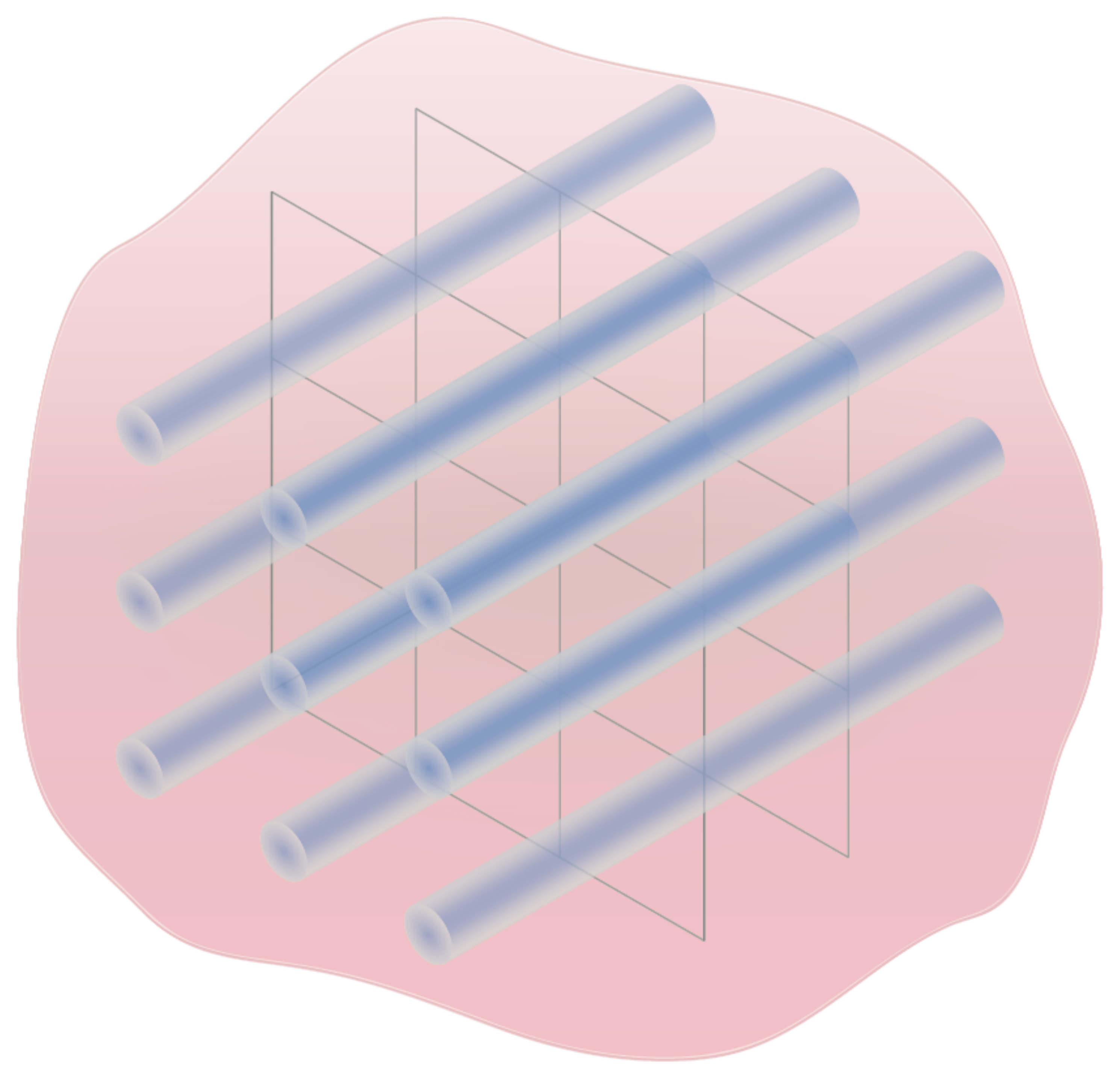}
\caption{\label{fig:fig1}Schematic representation of the system
studied in this work: A mixture of light fermions  and heavy bosons
loaded in an anisotropic optical lattice. As the bosons are assumed to be heavier,
the are confined to one dimension by the lattice potential.}
\end{figure}

  The interest in systems of interacting bosons and fermions has been a recurrent and central topic in
the study of the many-body problem. Many early studies were
concerned with dilute solutions of $^{3}$He in $^{4}$He (see
e.g.~\cite{baym}, for a review) as well  as with the problem of
electrons coupled to phonons in solids (see e.g.~\cite{AGD}). This
research led to the understanding of  important phenomena  such like
the  polaron and Cooper pairing~\cite{AGD}. More recently, these
concepts have reemerged in the context of ultracold atomic
gases~\cite{review_coldgases,mix3D}, where new types of interacting
Bose-Fermi mixtures have been experimentally
realized~\cite{Florence1,Florence2,Esslinger_BF,Sengstock_loc,Takahashi_mix,bfexp2,Bloch_selftrap,
Sengstock_selftrap2,bfexp1,Sengstock_selftrap,Inguscio_mixdim,Takahashi_lightheavymix}.
Indeed, such experiments with ultracold gases have made it possible
to  study and envision Bose-Fermi
systems~\cite{Das03,Pollet2008,Cazalilla03,DWmix1,DWmix2,Griffin07,Marchetti,screendiss1,screendiss2,
Rizzi08,Marchetti_oplatt,Tsai,Batrouni_SS1D,Kawakami,Troyer_cooper,Pieri,Xiwen,Hofstetter,Zwerger_BFSC,MalatsMix1}
that can exhibit very different properties from their
condensed-matter counterparts.

 Thus far, much research has  focused on understanding how interactions with the
bosonic component of the mixture influences the properties of the fermions and, in particular,
how the interactions mediated by the bosons  can possibly
induce  fermion superfluidity (see e.g.~\cite{review_coldgases,Zwerger_BFSC,Troyer_cooper} and references therein).
 The complementary problem, namely,   understanding how
the properties of bosons  are modified by their interaction with fermions  in a mixture
has only recently attracted interest, especially motivated by a series of ground-breaking
experiments with Bose-Fermi mixtures loaded in optical
lattices~\cite{Sengstock_loc,Sengstock_selftrap,Bloch_selftrap,Takahashi_mix,Sengstock_selftrap2}.

  Within this setup,  in recent years a number of  groups have addressed the problem of
how the addition of fermions to a Bose gas in an optical lattice affects the phase transition from superfluid to Mott
insulator in the latter~\cite{Esslinger_BF,Sengstock_selftrap,Bloch_selftrap,Takahashi_mix,Sengstock_selftrap2,screendiss1,screendiss2,KunYang_dissipation}.
Thus, experimental observations have been reported indicating that fermions effectively
decrease the quantum coherence of the bosons, thus making it easier for the latter to become
Mott insulating~\cite{Esslinger_BF,Sengstock_selftrap,Bloch_selftrap,Takahashi_mix,Sengstock_selftrap2}.
In the case of attractive interactions between bosons and fermions,
this effect has been explained by a `self-trapping'  effect: the bosons move in the lattice potential that
effectively  becomes deeper by the addition of  fermions to which the bosons are attracted.
This  `self-trapping' would have the opposite effect in the case of repulsive boson-fermion
interactions. However, in this case,  the two components were found  not mix in a deep lattice~\cite{Takahashi_mix}.

 In connection with the experiments referred to above, there has been some theoretical discussion on
other, perhaps more subtle,  effects of adding fermions to an interacting boson system~\cite{screendiss1,screendiss2}.  These effects concern the physics of polarons, where one particle (in this case, the bosons) is dressed by its interactions with a different species 
(the fermions).   Thus, the boson-boson interactions are screened, becoming much less repulsive. The bosons also undergo dissipative effects, which involve the creation of \emph{real} particle-hole pairs and other kinds of low-energy excitations in the Fermi gas.
Indeed, as discussed below, within a weak coupling approach  the `self-trapping effect' arises
at first order in the strength of the boson-fermion interactions, whereas the polaronic and dissipative effects arise at
 second order.

 Going beyond mean-field theory, Yang~\cite{KunYang_dissipation} has studied the effect of  the boson-fermions interaction on the superfluid (SF) to Mott insulator (MI) transition in a three dimensional Bose-Fermi mixture. He found that the properties of the transition at the particle-hole symmetric point (i.e.~at the tip of the `Mott lobes') is modified and becomes either a first order transition or a different (i.e.~not XY) second order transition.  However,  in this work we find that in a mixed dimensionality system, the universality class of the transition (2D XY) is not modified by the boson-fermion coupling for integer  filling.
The latter only introduces screening of the external periodic
potential and   the boson-boson interactions. Whereas the former
tends to make the system less (more) superfluid for attractive
(repulsive) boson-fermion interactions, the screening of the boson
interactions always favors superfluidity. On the other hand, for a
half-filled lattice, we find that the quantum phase transition from
the SF to the charge-density wave (CDW) phase is  modified by the
presence of the fermions. However, transition remains continuous and
belongs to the 2D XY universality class.

 The outline of this article is as follows. In the following section, we introduce the basic model of a Bose-Fermi mixture that
will be subsequently analyzed. There we also outline the derivation of its effective low-energy description.
In Sect.~\ref{sec:rg} we consider the effect of the Fermi gas on the Mott insulator to superfluid transition of a Bose gas
confined to one dimension~\cite{Haldane,giambook,review1D,pinning1d_exp}. The perturbative  renormalization group is used to analyze the low-energy properties of the effective low-energy model,  the effect of the fermions on the superfluid to insulator quantum phase transition is studied.
Depending on the lattice filling  and the boson-boson interactions,  the Mott insulator can be
stabilized at integer or half-integer filling, and the effects of the Fermi gas are very different on both transitions.
Thus, we have separated the discussion into two subsections,  \ref{sec:intf} and \ref{sec:halfint}.
Finally, in Sect.~\ref{sec:conclusions} the main conclusions of this work are summarized.

\section{Basic model}\label{sec:basicmodel}

\subsection{Hamiltonian}

 The system under study is an ultracold mixture of bosonic atoms (mass $m_B$) and
single-species fermionic atoms   (mass $m_F$) loaded in an  optical lattice (see Fig.~\ref{fig:fig1}).
The repulsive interaction between bosons is described by an interaction potential $v_{BB}(\mathbf{r}-\mathbf{r}')$. The latter
can be either the  Lee-Yang-Huang pseudo-potential, which accounts for the s-wave scattering of ultracold
atoms (as for alkali or alkaline earth atoms), or a  dipolar potential (as for Chromium or polar bosonic molecules). Furthermore,
fermions and bosons  are assumed to interact only via a short-range potential, which is also described by the Lee-Yang-Huang
pseudo-potential. Inter-fermion interactions are negligible because, by the Pauli principle, the dominant  scattering channel
for single-species fermions is  p-wave,
which,  away from resonances, is very weak at ultracold temperatures. The optical lattice potential  $U_{B(F)}(\mathbf{r})
=U^{B(F)}_{0\parallel}\sin^2{k_L x}+U^{F(B)}_{0\perp}(\sin^2{ k_L y}+\sin^2{ k_L z})$, where  $k_L =2\pi/\lambda_L$ and $\lambda_L$ is the laser wavelength. It is  further assumed that $U^{B}_{0\parallel}\ll U^B_{0\perp}$, that is, the bosons move in a strongly  anisotropic two-dimensional lattice. We further assume that the bosons are heavier (\emph{i.e.}~$m_F/m_B \ll 1$), which means  that
their motion along two directions (here $y$ and $z$) is strongly suppressed beyond zero-point motion,
thus effectively confining them to one dimension for at least the duration of the experiment.
However,  the fermions, being lighter, can hop in all three dimensions but  the large laser intensity required
to create the strong confining lattice potential for the bosons, which implies that $U^F_{0\perp} \ll U^F_{0\parallel}$,
the fermion dispersion will be anisotropic (see below Eq.~\ref{eq:fermdis}).  Thus,  the  Hamiltonian reads:
\begin{align}
\hat{H} &=\hat{H}_{B}+\hat{H}_{F}+\hat{H}_{BF}, \label{eq:basicham1}\\
\hat{H}_{B}&=\int d\mathbf{r}\Big[\frac{\hbar^2}{2m_{B}}\nabla\hat{\Psi}_B^{\dag}(\mathbf{r})\nabla\hat{\Psi}_B(\mathbf{r})  + U_B(\mathbf{r})
\hat{\rho}_{B}(\mathbf{r}) \nonumber\\
&\quad   +\frac{1}{2} \int d\mathbf{r}^{\prime}  \: \hat{\rho}_{B}(\mathbf{r})V_{BB}(\mathbf{r}-\mathbf{r})\hat{\rho}_{B}(\mathbf{r'})\Big],\label{eq:basicham2}\\
\hat{H}_{F} &=\int d\mathbf{r}\Big[\frac{\hbar^2}{2m_{F}}\nabla\hat{\Psi}_F^{\dag}(\mathbf{r})\nabla\hat{\Psi}_F(\mathbf{r}) +
U_F(\mathbf{r}) \hat{\rho}_{F}(\mathbf{r})\Big],\label{eq:basicham3}\\
\hat{H}_{BF}&=g_{BF}\int d\mathbf{r}\, \hat{\rho}_{B}(\mathbf{r})\hat{\rho}_{F}(\mathbf{r}), \label{eq:basicham4}
\end{align}
where  $\hat{\Psi}_{B(F)}(\mathbf{r})$ is the boson (fermion) field
operator, which obeys $\left[\hat{\Psi}^{\dag}_B(\mathbf{r}),
\hat{\Psi}_{B}(\mathbf{r}') \right] = \delta(\mathbf{r}-\mathbf{r})$
($\{ \hat{\Psi}^{\dag}_F(\mathbf{r}), \hat{\Psi}_{F}(\mathbf{r}') \}
= \delta(\mathbf{r}-\mathbf{r})$) (anti-)commuting otherwise;
$\hat{\rho}_{B(F)}(\mathbf{r})=\hat{\Psi}^{\dag}_{B(F)}(\mathbf{r})\hat{\Psi}_{B(F)}(\mathbf{r})$
is the boson (fermion) density operator and $N_{B(F)}  = \int
d\mathbf{r}\: \hat{\rho}_{B(F)}(\mathbf{r})$ the boson (fermion)
number operator. The  boson-fermion interaction is parametrized by
the coupling $g_{BF}=2\pi\hbar^2a_{BF}/M_{BF}$, where
$M_{BF}=m_{B}m_{F}/(m_{B}+m_{F})$ is the reduced mass and $a_{BF}$
is the s-wave scattering length. Since we are interested in the
ground state phase diagram in the thermodynamic limit of the above system, we
have neglected the harmonic trapping potential, which is also present in the
experiments. Note that an implicit assumption of our analysis below
is that the bosons and fermions are mixed. For short range
interactions between the bosons (i.e. for $V_{BB}(\mathbf{r}) =
g_{BB}\delta(\mathbf{r})$) the problem of the bosons and fermions forming a
uniform mixture in the lattice geometry studied here has been
previously considered in Ref.~\cite{MalatsMix1}. One conclusion of this
work is that the uniform mixed phase in this Bose-Fermi system is
always stable provided the density of bosons and fermions is
sufficiently high, for both attractive and repulsive interactions
(see Ref.~\cite{MalatsMix1} for  further details).

  The  Hamiltonian introduced in equations~(\ref{eq:basicham1}, \ref{eq:basicham2}, \ref{eq:basicham3}), and \eqref{eq:basicham4}
contains too much information about energy scales in which we are not interested. Since our goal is to analyze the ground
state and low-lying excitations of the system, we next derive an effective Hamiltonian that is much more appropriate
to this end. The first step  is to project the Bose and Fermi fields onto the lowest Bloch band of the
lattice potential. Thus,  we expand $\hat{\Psi}_B(\mathbf{r}) \simeq \sum_{\mathbf{R}} w_0(\mathbf{r}_{\perp} - \mathbf{R}) \:
\hat{\Psi}_{B{\mathbf{R}}}(x)$ where $w_0(\mathbf{r}_{\perp} - \mathbf{R})$ are the Wannier orbitals describing particles
localized round the site $\mathbf{R} = \frac{1}{2}(m, n)\lambda_L$ of a 2D (square) lattice.
For the fermions, $\hat{\Psi}_F(\mathbf{r}) \simeq \sum_{\mathbf{k}} \varphi_{\mathbf{k}}(\mathbf{r})\, \hat{f}_{\mathbf{k}}$, where
$\varphi_{\mathbf{k}}(\mathbf{r})$ are the Bloch states of the lowest band.
Note the differences in treatment of the Bose and Fermi fields, which reflects their differences in mobility introduced
by the conditions discussed above. Hence, upon neglecting terms coupling different lattice sites, the bosons
are described by
\begin{align}
\hat{H}_{B} &=  \sum_{\mathbf{R}} \int dx \,  \left[  \frac{\hbar^2}{2m_B} \left|\partial_x \hat{\Psi}_{B\mathbf{R}}(x) \right|^2
 +   U_{B\|}(x)  \hat{\rho}_{B\mathbf{R}}(x) \right] \nonumber  \\
& \quad +  \frac{1}{2} \sum_{\mathbf{R}}\int dx dx'\,  V_{BB}(x-x')\hat{\rho}_{B\mathbf{R}}(x) \hat{\rho}_{B\mathbf{R}}(x^{\prime}).
\end{align}
However, the fermions are described by:
\begin{equation}
\hat{H}_F= \sum_{\mathbf{k}} \epsilon(\mathbf{k}) \: \hat{f}^{\dag}_{\mathbf{k}} \hat{f}_{\mathbf{k}},\label{eq:fermdis}
\end{equation}
where the sum is over $\mathbf{k}$ belonging to the first Brioullin zone and $\epsilon(\mathbf{k}) = \epsilon_{\parallel}(k) +
\epsilon_{\perp}(\mathbf{k}_{\perp}) \simeq \frac{\hbar^2 k^2}{2m^*_F}  -2 t_{\perp} \left( \cos k_y b_0 + \cos k_z b_0  \right)$, where
$b_0 = \frac{\pi}{k_L}$ is the lattice parameter,  and we have assumed that the periodic potential along the $x$ direction is so weak that effectively amounts to a renormalization of the fermion mass.   Finally, the boson-fermion interactions are described by:
\begin{equation}
\hat{H}_{BF}  = g_{BF} \sum_{\bf R} \int \: d\mathbf{r} \, \left|w_0(\mathbf{r}_{\perp} - \mathbf{R})  \right|^2 \,
\hat{\rho}_{B\mathbf{R}}(x) \hat{\rho}_F(\mathbf{r}),
\end{equation}
where $\mathbf{r}= (x,y,z) = (x,\mathbf{r}_{\perp})$. In the above expressionm we have approximated the
boson density operator $\hat{\rho}_B(\mathbf{r}) = \hat{\rho}_B(x,\mathbf{r}_{\perp})
\simeq \sum_{\mathbf{R}} \left|w_0(\mathbf{r}_{\perp} - \mathbf{R})  \right|^2 \hat{\rho}_{B\mathbf{R}}(x) $.

\subsection{Integrating out the fermions}\label{subsect:intferm}

The total Hamiltonian obtained upon projection
onto the lowest Bloch band $H = H_{B} + H_{F}  + H_{BF}$ is still too complicated to solve. Since we
are mainly interested on the low-temperature properties of the heavier bosons, which are much slower,
a first step towards understanding the latter is integrating  out the fermion degrees of freedom.
To this end, we rely on the path integral representation of the partition function  $Z = \mathrm{Tr} \, e^{-\beta \left[ H  -\mu_B N_B -  \mu_F N_F \right]}$  for  the Hamiltonian, $H = H_B + H_F + H_{FB}$, which allows us to write:
\begin{equation}
Z  = \int \left[d\bar{\psi}_B d\psi_B d\bar{\psi}_F d\psi_F \right]  \, e^{-S[\bar{\psi}_B,\psi_B,\bar{\psi}_F,\psi_F]},
\end{equation}
where
\begin{align}
S &= S_B+ S_F + S_{BF},\notag\\
S_B &= \sum_{\mathbf{R}}\int dx \int^{\hbar\beta}_0 \,  d\tau \,
\bar{\psi}_{B\mathbf{R}}(x,\tau) \partial_{\tau} \psi_{B\mathbf{R}}(x,\tau) \notag\\
  &- \frac{\mu_B}{\hbar}  \sum_{\mathbf{R}}\int dx
 \int^{\hbar\beta}_0 d\tau |\psi_{B\mathbf{R}}(x,\tau)|^2   \notag \\
& \quad
 + \int^{\hbar\beta}_0 \frac{d\tau}{\hbar}  \:  H_B(\tau),\\
S_F&= \sum_{\mathbf{k}} \int^{\hbar\beta}_0 d\tau
\bar{f}(x,\mathbf{k}) \left[ \partial_{\tau} f(\mathbf{k},\tau) - \frac{\mu_F}{\hbar} f(\mathbf{k},\tau) \right]  \notag\\
 &\quad + \frac{1}{\hbar} \int^{\hbar\beta}_0 d\tau \:  H_F(\tau),\\
 S_{BF} &= \frac{1}{\hbar} \int d\tau \, H_{BF}(\tau).
\end{align}
where $\beta = (k_B T)^{-1}$ is the inverse of absolute temperature and $k_B$ is Boltzmann's constant.
Thus, the effective action for the bosons is defined by the following equation:
\begin{align}
e^{-S_\mathrm{eff}[\bar{\psi}_B,\psi_B]} &= \int  \left[d\bar{f}df \right] \,   e^{-S_B - S_F- S_{BF}}\nonumber \\
 &= Z^0_F \, e^{-S_B}  \langle e^{- S_{BF}}\rangle_F,
\end{align}
where $\langle\ldots \rangle_F = \mathrm{Tr}\:  \hat{\rho}_F \ldots$ and $\hat{\rho}_F =  Z^{-1}_F\:
e^{-\beta (H_F-\mu N_F)} $, being $Z_F = \mathrm{Tr}\: e^{-\beta (H_F-\mu N_F)}$ the non-interacting fermion partition
function. To make further progress, we shall assume that the interaction between the bosons and the fermions
is pertubatively small. Therefore, the above functional integral can be performed using the cumulant expansion,
which yields:
\begin{equation}
\langle e^{-S_{BF}} \rangle_F = e^{-\langle S_{BF}\rangle + \frac{1}{2} \langle S^2_{BF} - \langle S_{BF}\rangle^2\rangle+\cdots}
\end{equation}
The leading term is
\begin{align}
\langle S_{BF} \rangle_F  &= \frac{g_{BF}}{\hbar}  \sum_{\mathbf{R}} \int^{\hbar\beta}_0 d\tau \int d\mathbf{r} |w_0(\mathbf{r}_{\perp}-\mathbf{R})|^2 \notag\\ &\quad\quad \times \rho_{B\mathbf{R}}(x,\tau) \rho^0_F(\mathbf{r}),\label{eq:sbf1}
\end{align}
where $\rho^0_F(\mathbf{r}) = \langle \rho_F(\mathbf{r},\tau)\rangle_F$ is the equilibrium density of the
Fermi gas (in the absence of the bosons). Since $\rho^0_F(\mathbf{r})$ is periodic, \eqref{eq:sbf1} amounts to a
correction to the periodic potential that the boson gas undergoes. The correction has the same sign
as the coupling $g_{BF}$, which means  that e.g.~for attractive boson-fermion interactions, the effective
potential seen by the bosons is deepened by its (mean-field) interaction with the fermions. This
effect has been termed `self-trapping' and has been studied both theoretically~\cite{Sengstock_selftrap2} and
experimentally~\cite{Sengstock_selftrap,Bloch_selftrap,Takahashi_mix}.  We shall not study it any further here. Instead, we focus on the second order  term, which leads to much more interesting physics. Neglecting the coupling between different
sites $\mathbf{R}$ (i.e.~terms where $\mathbf{R}^{\prime}\neq \mathbf{R}$) yields:
\begin{align}
- \frac{1}{2}\langle S^2_{BF} - \langle S_{BF}\rangle^2 \rangle   =  \frac{g^2_{BF}}{2\hbar} \sum_{\mathbf{R}}
\int d\mathbf{r} d\tau d\mathbf{r}^{\prime}  d\tau^{\prime} \rho_{B\mathbf{R}}(x,\tau)  \notag\\
\times \chi_F(x-x^{\prime},\tau-\tau^{\prime}) \rho_{B\mathbf{R}}(x^{\prime},\tau^{\prime}),\quad\quad\quad
\label{eq:sbfeff}
\end{align}
where $\mathbf{r} = (x,\mathbf{r}_{\perp})$, $\mathbf{r}^{\prime} =
(x^{\prime},\mathbf{r}^{\prime}_{\perp})$. After defining $F_0(\mathbf{r}_{\perp},
\mathbf{r}_{\perp}) = |w_0(\mathbf{r}_{\perp})
w_0(\mathbf{r}^{\prime}_{\perp})|^2$, we introduce
\begin{align}
\chi_F(x,\tau) &= \int d\mathbf{r}_{\perp} d\mathbf{r}^{\prime}_{\perp}\,
F_{0}(\mathbf{r}_{\perp},\mathbf{r}^{\prime}_{\perp})  \chi_F(\mathbf{r},\mathbf{r}^{\prime},\tau), \notag \\
\chi_F(\mathbf{r},\mathbf{r}^{\prime},\tau) &=  - \frac{1}{\hbar}
\langle \delta \rho_F(\mathbf{r},\tau) \delta \rho_F (\mathbf{r}^{\prime},0) \rangle_F.
\end{align}
Thus, up to  $O(g^2_{BF})$,
we obtain the following effective action for the bosons:
\begin{align}
S_{\mathrm{eff}}[\psi^*_B,\psi_B] &= \sum_R S_{\mathrm{eff},\mathbf{R}} \notag\\
S_{\mathrm{eff},\mathbf{R}} &= \int^{\hbar\beta}_0 d\tau \int dx\:
\psi^*_{B\mathbf{R}}(x,\tau) \partial_{\tau} \psi_{B\mathbf{R}}(x,\tau) \notag\\
  &+  \int^{\hbar\beta}_0 d\tau \:  \int dx \frac{\hbar}{2m_B} |\partial_x \psi_{B\mathbf{R}}(x,\tau)|^2  \notag\\
  &  +      \int^{\hbar\beta}_0  \frac{d\tau}{\hbar} \int dx \left[ \tilde{U}_{B\parallel}(x) - \mu_B \right] |\psi_{B\mathbf{R}}(x,\tau)|^2   \notag\\
 &+  \frac{1}{2}  \int^{\hbar\beta}_0 \frac{d\tau}{\hbar} \, \int dx dx^{\prime}
  |\psi_{B\mathbf{R}}(x,\tau)|^2
  \notag\\
 &\quad\quad\quad  \times  V_{BB}(x-x^{\prime}) |\psi_{B\mathbf{R}}(x^{\prime},\tau)|^2\notag\\
  &+ \frac{g^2_{BF}}{2\hbar}
\int dx d\tau d x^{\prime}  d\tau^{\prime} |\psi_{B\mathbf{R}}(x,\tau)|^2  \notag\\
& \times \chi_F(x, x^{\prime},\tau-\tau^{\prime}) |\psi_{B\mathbf{R}}(x^{\prime},\tau^{\prime})|^2,
\label{eq:effact}
   \end{align}
where $\tilde{U}_{B\parallel}(x)  = U_{B\parallel}(x)  + g_{BF} \int d\mathbf{r}_{\perp} \: |w_0(\mathbf{r}_{\perp})|^2 \: \rho^0_F(x,\mathbf{r}_{\perp})$. Note that we have thus reduced the problem to a set of one dimensional systems independently coupled to a fermionic bath.
Therefore, in what follows we shall drop the lattice index $\mathbf{R}$ and study the phase diagram of a generic 1D system
coupled to the fermionic bath.

 However, one important caveat is in order when considering the applicability of the  effective action, Eq.~\eqref{eq:effact}. In what follows, we will not treat the bosons and the fermions on equal footing. Such a treatment would require to also account for  the effect of the bosons on  the fermionic component of the mixtures, which may modify the density response $\chi_F(\mathbf{r},\mathbf{r}^{\prime},\tau)$. Nevertheless, below we shall assume  that $\chi_F(\mathbf{r},\mathbf{r}^{\prime},\tau)$  is well described by the non-interacting
limit where we take $g_{BF}  =0$. Indeed, this assumption
 is qualitatively correct as long as the Fermi component of the mixture remains a Fermi liquid, which is reasonable given
 that the fermions are much lighter, interact with the bosons weakly, and therefore their energy is dominated by the kinetic
 energy.  However, strictly speaking the bosons will mediate  effective fermion-fermion interactions, which, at sufficiently low temperature, lead to a pairing  instability of the Fermi gas. Since the gas contains a single species of fermions, such  a paring instability takes
 place in a high angular momentum wave (most likely,  p-wave) and  at relatively low temperatures compared to the Fermi energy
 $\mu_F$. Given that present cooling techniques in optical lattices cannot reach temperatures below a few percent of $\mu_F$,
 we can safely neglect this possibility. Other  instabilities that can gap the fermion spectrum, such as a
 charge density wave,  occur at particular values of the lattice filling and/or lattice parameters and we will also
 neglect them in what follows.

\begin{figure}[b]
\includegraphics[height=0.30\textwidth]{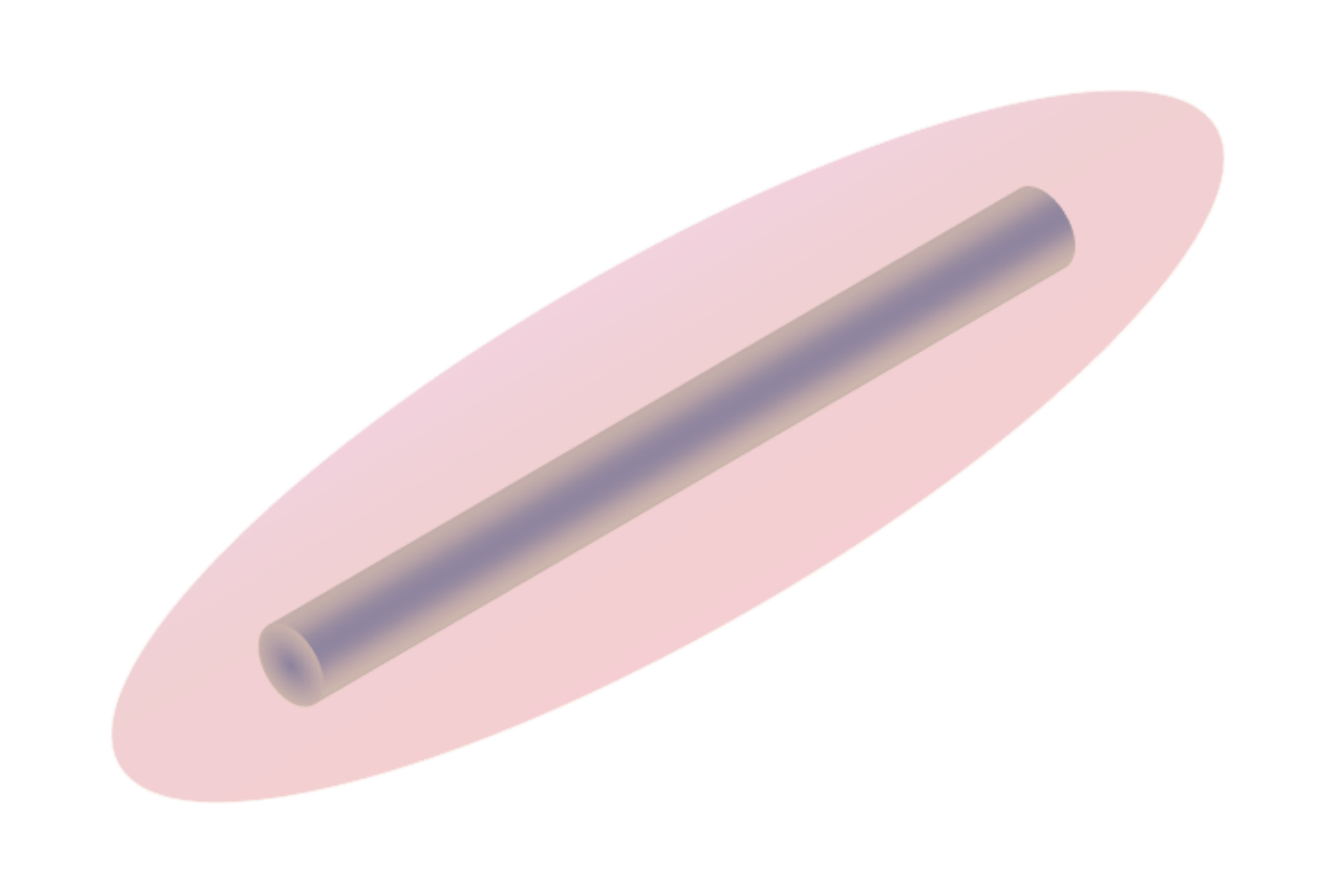}
\caption{\label{fig:fig2} In the limit of strong one-dimensional (1D) confinement, the inter-tube coupling can be neglected and
 the system of Fig.~\ref{fig:fig1} can be mapped to  a single 1D tube immersed in a Fermi gas. The Fermi gas has
 two important effects: it introduces screening of the boson-boson interactions and it behaves as a dissipative bath
 that introduces quantum dissipation.  Assuming that
a periodic potential is applied longitudinally to the tube(s), which drives a transition from the superfluid
to a Mott insulating phase, we study  the modification of the phase diagram due to these effects.}
\end{figure}

\subsection{Low-energy effective theory}

 In order to deal with the effective boson model in Eq.~(\ref{eq:effact}),
we shall use the method of bosonization~\cite{review1D,Gogolin}.
Thus, we first integrate the high-energy density and phase fluctuations of the bosons, and introduce
two collective fields, $\theta(x)$,  and $\phi(x)$ describing the phase fluctuations in each 1D system.
In terms of these fields, the Bose field and density operators read:
\begin{align}
\Psi_{B}(x) &\simeq  \mathcal{A} \: \rho^{1/2}_0 \:  e^{i\theta(x)}, \\
\rho_{B}(x) &= \Psi^{\dag}_{B}(x) \Psi_B(x)\simeq \rho_0 + \frac{1}{\pi} \partial_x \phi(x)  \nonumber \\
 & + \rho_0 \sum_{m > 0} \mathcal{B}_m  \cos 2m \left( \phi(x) + k^{B}_F x\right),\label{eq:bden}
\end{align}
where $\rho_0 = N_{B}/(ML)$ is the linear density of bosons in each of the $M$ 1D systems of length $L$ of the lattice
and $k^B_F = \pi \rho_0$.  The amplitudes $\mathcal{A}$ and $\mathcal{B}_m$ depend on the microscopic details of the model and
can not be obtained using bosonization. Using the above expressions and retaining only the most relevant operators in the
renormalization-group sense yields~\cite{review1D}:
\begin{align}
S_B[\phi] &= S_{0}[\phi] + S_{u}[\phi],\label{eq:sb} \\
S_{0}[\phi]  &= \frac{1}{2\pi K} \int dx \left[ \frac{1}{v} \left(\partial_{\tau}\phi \right)^2 +
v \left(\partial_x \phi \right)^2\right],\label{eq:sb0}\\
S_{u}[\phi] &=- \frac{g_{u}}{\pi a^2_0} \int d\tau dx \, \cos \left( 2p \phi(x,\tau) + x \delta_p \right),\label{eq:su}
\end{align}
where we have introduced the following notation: $v$
is the sound velocity of the 1D Bose gas whereas $vK = v^B_F = \hbar k^B_F/m_B$ and $K/v$
is proportional to the system compressibility; $a_0\approx \hbar v/\mu_B$ is a short-distance
cut-off~\cite{review1D}.

   The bare \emph{dimensionless} coupling of the term describing the periodic potential in the bosonization
  language is  $g_u = \tilde{U}_{0\parallel} (\mathcal{B}_p \rho_0 a^2_0)/2\hbar v$.
The cosine term with $p = 1$ describes the effect of the potential
in the case of integer filling of the lattice,  $\delta_{p=1} = 2
(k_L - k^{B}_F)$ being a measure of the incommensurability of the
system. However, near half filling,  we must consider the $p = 2$
term with $\delta_{p=2} = 2 k_L - 4 k^{B}_F$ as a measure of the
incommensurability. In the half-filled case, the above effective
Hamiltonian describes the transition from a Tomonaga-Luttinger
liquid (TLL) to a fractional Mott insulating state  which is also
known as a charge density wave (CDW).  The  stability of the CDW
state requires  smaller values of the  Luttinger parameter
$K$~\cite{review1D} than those that are achievable in the
Lieb-Liniger model~\cite{Lieb63} describing bosons interacting via a
short range potential in a 1D wave-guide~\cite{Olshanii98}, for
which the minimum value of $K$ is one~\cite{review1D}. Smaller
values  of $K$ are accessible when either the bosons posses a
dipolar moment~\cite{Orignac_dipolar,review1D}  or in the so-called
super-Tonks regime~\cite{supertonks,review1D,supertonks_exp} .

\begin{figure}[ht]
\includegraphics[height=0.30\textwidth]{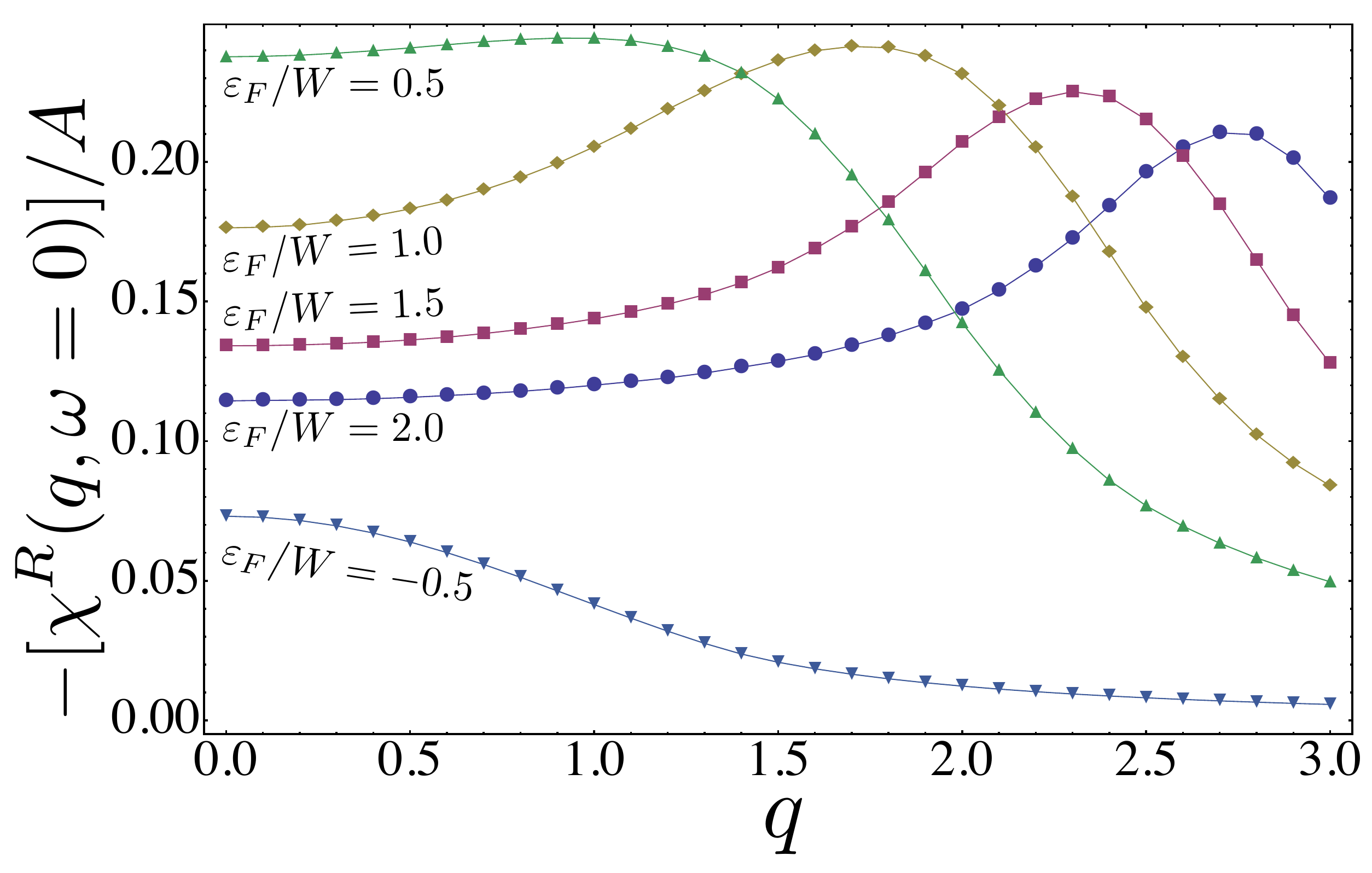}
\caption{\label{fig:rechi} The static response function $\chi^{R}_F(q,\omega=0)/A$  of the fermions 
for different values of the Fermi energy $\epsilon_{F}=2.0,1.5,1.0,0.5,-0.5$ in units where $\frac{\hbar^2}{2m}=1$ and $W=4 t_{\perp} = 1$. See Appendix~\ref{app:response} for  details of the calculation.}
\end{figure}

\begin{figure}[hb]
\includegraphics[height=0.30\textwidth]{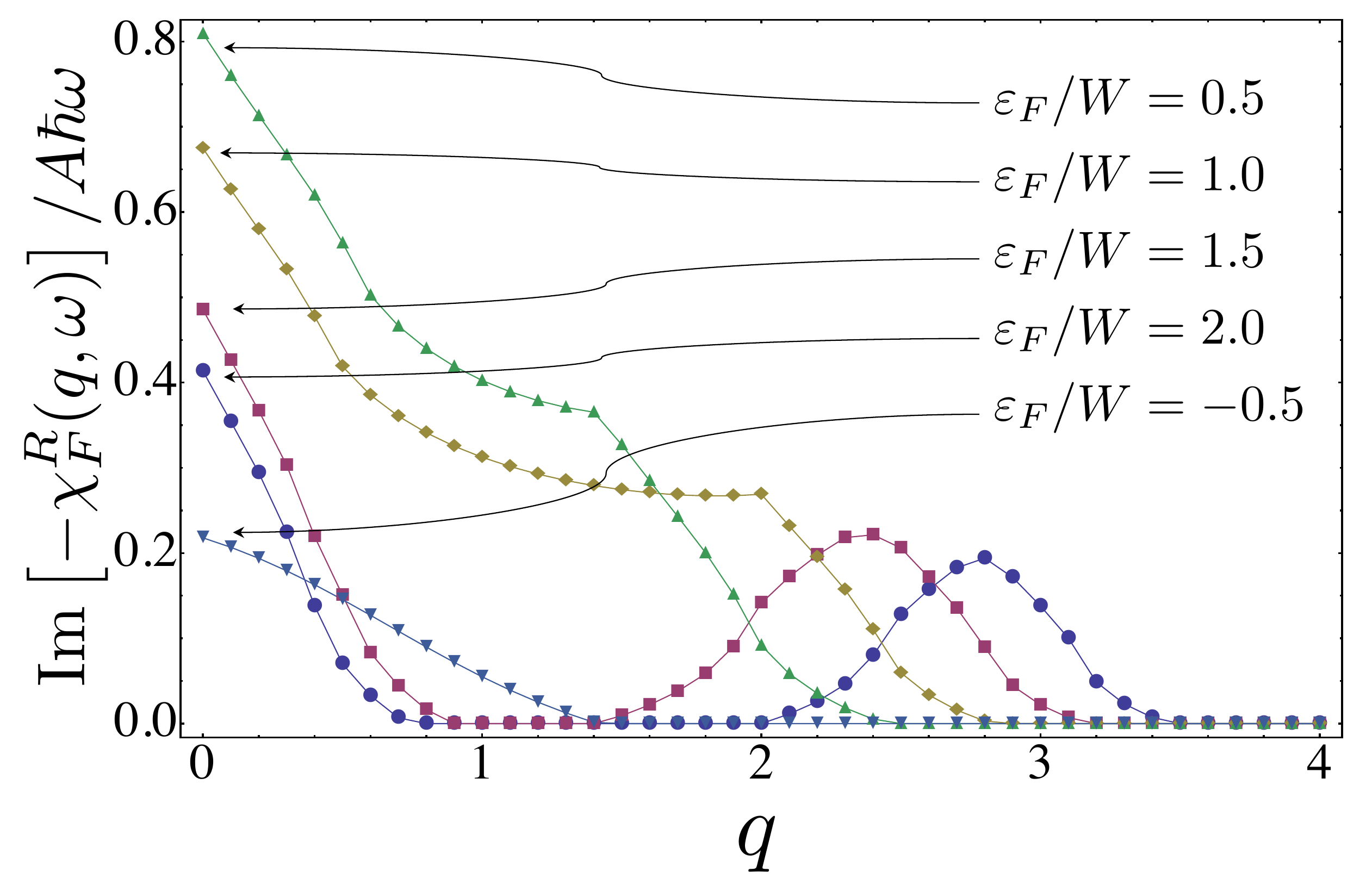}
\caption{\label{fig:imchi} Imaginary part of the fermion response function divided by the excitation frequency, $\omega$, 
for $\omega \to 0^+$,  for different values of the Fermi energy $\epsilon_{F}=2.0,1.5,1.0,0.5,-0.5$. 
Units where $\frac{\hbar^2}{2m_{F}}=1$ and $W = 4t_{\perp}=1$ have been used. See Appendix~\ref{app:response} for  details of the calculation.}
\end{figure}

The above action, Eq.~\eqref{eq:sb}, provides an effective description of the low-temperature properties of the boson system
which includes (through the renormalization of the potential $U_{B\parallel} \to \tilde{U}_{B\parallel}$) the effect of
the Fermi gas at the mean-field level. The dynamical effect of the fermions on the bosons
is taken  into account, to leading order in $g_{BF}$, by the last term
in Eq.~\eqref{eq:effact}. However, since the dynamics of the (heavier) bosons
described by \eqref{eq:sb0} is much slower than the lighter fermions, some further simplifications of
\eqref{eq:sbfeff} are possible. First, we  note  (see Appendix~\ref{app:response}) that, at $T = 0$, the fermion density correlation
function introduced above, $\chi_F(x,x^{\prime},\tau)$ can be written as follows:
\begin{equation}
\chi_F(x - x^{\prime},\tau) = \int^{+\infty}_0 \frac{d\omega}{\pi} \, e^{-\omega |\tau|} \, \mathrm{Im} \chi^R_F(x - x^{\prime},\omega),
\end{equation}
where $\chi^R_F(x-x^{\prime},\omega)$ is the retarded version of the same correlation function.
We have also assumed, consistently with what was stated above, that the effect of the periodic potential
can be neglected as far as the calculation of $\chi_F(x-x^{\prime},\omega) \simeq \chi_F(x-x^{\prime},\omega)$.
The above expression allows us
to treat separately the high frequency density fluctuations from the low-frequency fluctuations of the fermionic
gas. This can be done by introducing the following response functions:
\begin{align}
\chi^{<}_F(x,\tau) &= \int^{+\infty}_0  \frac{d\omega}{\pi} \,  g(\omega) e^{-\omega |\tau|} \, \mathrm{Im} \chi^R_F(x,\omega), \label{eq:chim}\\
\chi^{>}_F(x,\tau) &= \int^{+\infty}_0  \frac{d\omega}{\pi} \, g_c(\omega) e^{-\omega |\tau|} \, \mathrm{Im} \chi^R_F(x,\omega),
\end{align}
where $g(\omega)$  is a frequency cut-off function, which can be chosen in various ways as the result will be largely
independent of this function; $g_c(\omega) = 1 - g(\omega)$. Below we use $g(\omega) = e^{-\omega \tau_c}$, where $\tau_c
\ll \max\{\frac{\hbar}{\mu_F}, \frac{\hbar}{\mu_B}\}$. The cut-off frequency $\simeq \frac{\hbar}{\tau_c}$ is chosen such
that the high-frequency density fluctuations of the Fermi gas  can adapt instantaneously to the (slow) dynamics of the boson density
fluctuations described by $\rho_B(x,\tau)$ (cf.  Eq.~\ref{eq:bden}). Thus,
\begin{align}
& \int dx d\tau dx^{\prime} d\tau^{\prime} \: \rho_B(x,\tau) \chi^{>}_F(x-x^{\prime},\tau-\tau^{\prime})
\rho_B(x^{\prime},\tau^{\prime}) \notag\\
&= \int dx dx^{\prime} dt d\tau \: \rho_B(x,\tau + \frac{t}{2}) \chi^{>}_F(x-x^{\prime},t) \rho_B(x,\tau - \frac{t}{2}) \notag\\
&\simeq  \int dx dx^{\prime} d\tau \rho_B(x,\tau) \chi^{>}_F(x-x^{\prime},\omega=0) \rho_B(x^{\prime},\tau)\quad\\
&=  \int dx dx^{\prime} d\tau \rho_B(x,\tau) \chi_F(x-x^{\prime},\omega=0) \rho_B(x^{\prime},\tau)  \notag\\
 & \quad- \int dx dx^{\prime} d\tau \rho_B(x,\tau) \chi^{<}_F(x-x^{\prime},\omega=0) \rho_B(x^{\prime},\tau),
\end{align}
where $\chi^{>}_F(x-x^{\prime},\omega=0) = \int dt \, \chi^{>}_F(x-x^{\prime},t)$ and similar definitions
for $\chi^{<}_F(x-x^{\prime},\omega=0)$ and $\chi_F(x-x^{\prime},\omega=0)$. Therefore,
the effective action describing the interactions between the bosons mediated by the fermi gas
takes the form:
\begin{align}
S_{\mathrm{eff},BF} &= \frac{g^2_{BF}}{2\hbar} \int dx dx^{\prime} d\tau \: \rho_B(x,\tau) \chi_F(x-x^{\prime},\omega=0)
\notag\\
\quad\quad &\times \rho_B(x^{\prime},\tau) +  \frac{g^2_{BF}}{2\hbar} \int dx dx^{\prime} d\tau d\tau^{\prime} \rho_B(x,\tau) \notag\\
 &\quad\quad\quad\times \Gamma(x - x^{\prime},\tau-\tau^{\prime}) \rho_B(x^{\prime},\tau^{\prime}), \label{eq:sbff}
\end{align}
where the \emph{dissipative} kernel $\Gamma(x, x^{\prime}, \tau)$ is defined as:
\begin{equation}
\Gamma(x- x^{\prime},\tau) = \chi^{<}_F(x,\tau) - \chi^{<}_F(x-x^{\prime},\omega=0) \delta(\tau).
\end{equation}
Note that, by definition, $\int d\tau \, \Gamma(x-x^{\prime},\tau) = 0$. This
kernel can be evaluated as follows. Since we assume the Fermi component of the mixture to be a
Fermi liquid, we note that for the latter $-\mathrm{Im} \: \chi^R_F(x-x^{\prime},\omega)
\propto \omega$ for $\omega \ll |\mu_F|$~\cite{baym}.
In the present system, the small $\omega$ limit of this function is obtained
explicitly in Appendix~\ref{app:response} at $T = 0$. It can be written as
\begin{equation}
\mathrm{Im} \, \chi_F(x- x^{\prime},\omega\ll  \frac{\hbar}{\tau_c}) = - \pi D(x-x^{\prime}) \omega.
\end{equation}
where $D(x)$ is a positive function of $x$ which is computed in Appendix~\ref{app:response}.
Introducing this expression into \eqref{eq:chim}, yields:
 \begin{equation}
 \Gamma(x-x^{\prime}, \tau)  =  - \frac{D(x-x^{\prime})}{(|\tau| +  \tau_c)^2}  \label{eq:gammaqt}
 \end{equation}
at $T = 0$.  Introducing the above expression into Eq.~\eqref{eq:sbff}, we arrive at:
\begin{align}
S_{\mathrm{eff},BF} &= \frac{g^2_{BF}}{2\hbar} \int dx dx^{\prime} d\tau \: \rho_B(x,\tau) \chi_F(x-x^{\prime},\omega=0)
\notag\\
\quad\quad &\times \rho_B(x^{\prime},\tau) - \frac{g^2_{BF}}{2\hbar} \int dx dx^{\prime} d\tau d\tau^{\prime} \rho_B(x,\tau) \notag\\
 &\quad \times  \frac{D(x-x^{\prime})}{(|\tau-\tau^{\prime}|+\tau_c)^2} \rho_B(x^{\prime},\tau^{\prime}), \label{eq:sbf}
\end{align}
The results of the model calculation described in Appendix~\ref{app:response} for the functions $ D(q)/\hbar = - \mathrm{Im} \:
\chi^R_F(q,\omega)$ (for $\omega \ll \frac{\hbar}{\tau_c}$ and the static response function $\chi^R_F(q,\omega = 0)$ are displayed in
Figs.~\ref{fig:imchi} and \ref{fig:rechi}. It can be seen that both functions are  rather smooth (i.e. non-singular) functions of the longitudinal
wavevector $q$. This assumption will prove important below. Furthermore, for certain values of the lattice filling, which determines
the Fermi energy $\epsilon_F$, see Appendix~\ref{app:response},  $D(q)$ can be  made negligible or zero for wide ranges of
the wavevector $q$. This opens the possibility  of tuning the strength of the dissipative effects by simply changing the
fermion density. Note, however, that by strongly reducing the fermion density, the stability of the mixture may be
jeopardized~\cite{MalatsMix1}.


Thus we see that the boson interaction mediated by the Fermi gas consists, at low frequencies, of an
instantaneous part (which stems for high frequency density fluctuations of the Fermi gas) and a dissipative
part, which takes the form of a retarded $\sim \frac{1}{\tau^2}$ interaction. The latter stems from the
excitation by the motion of the bosons of \emph{real} low-energy particle-hole pairs, which
in a Fermi liquid yield the linear-$\omega$ behavior of the density response function (i.e. Landau damping). As
discussed above, the instantaneous part of the interaction can related to the static density response of the
Fermi gas and leads to a renormalization of the sound velocity $v$ and  Luttinger parameter $K$ describing the low-temperature properties of 1D boson system. The renormalized parameters obey:
\begin{align}
\frac{v(g_{BF})}{K(g_{BF})} &= \frac{v(g_{BF} = 0)}{K(g_{BF} =0)} \notag\\
&+ 2 \frac{g^2_{BF}}{\hbar} \chi_F(q = 0, \omega =0),
\label{eq:renparam}
\end{align}
Furthermore, since the fermion-induced interaction is a density-density interaction (cf. first term in
Eq.~\ref{eq:sbf}), we have that~\cite{review1D}:
\begin{equation}
v(g_{BF}) K(g_{BF})= v(g_{BF} = 0) K(g_{BF} =0).
\end{equation}
These equations describe, to lowest order in $g_{BF}$,
the \emph{screening}  of the boson-boson interaction by the fermion gas,
which leads to corrections to the parameters $K$ and $v$ in Eq.~\eqref{eq:sb0},
which  depend only on the boson-boson interaction.

Using the bosonization formula~\eqref{eq:bden}, we obtain the representation of the
dissipative action in terms of the density field $\phi(x,\tau)$:
\begin{align}
\tilde{S}_{D} &= S^f_{D} + S^b_{D} \label{eq:sbd}\\
S^f_{D} &= -\frac{\tilde{g}_{f D}}{\pi^2} \int dx d\tau d\tau^{\prime}  \frac{\partial_x \phi(x,\tau) \partial_x \phi(x,\tau^{\prime})}{
(|\tau-\tau^{\prime}| + \tau_c)^2}\label{eq:sdissf},\\
S^b_{D} &= - \frac{g_{b D}}{a_0}  \int dx d\tau d\tau^{\prime} \, \frac{\cos 2 \left[ \phi(x,\tau) -
\phi(x,\tau')\right]}{(|\tau-\tau^{\prime}| + \tau_c)^2}. \label{eq:sdissb}
\end{align}
In the derivation of the above  perturbations to the Gaussian action, Eq.~\eqref{eq:sb0}, we have
retained only terms whose integrands are not oscillatory and are the leading terms in a gradient
expansion. However, in  the case of a half-filled lattice,  the following term:
\begin{equation}
S^u_{D} = - \frac{g_{u D}}{a_0}  \int dx d\tau d\tau^{\prime} \, \frac{\cos 2 \left[ \phi(x,\tau) + \phi(x,\tau')\right]}{(|\tau-\tau^{\prime}| + \tau_c)^2}, \label{eq:sdissu}
\end{equation}
must be also taken into account.  This dissipative \emph{umklapp} interaction arises from the periodicity of
the boson system, for which which at half-filling $4 k^B_F = \frac{2\pi}{b_0}$, is a reciprocal lattice wave number.
In this regard, we must recall that, in a periodic system, the (lattice) momentum along the $x$ direction is
conserved modulo a reciprocal lattice wave number. Note that this term will be also generated by the renormalization
group flow from product of the  $S_{u}$  (cf. Eq.~\ref{eq:su}) and $S^b_{D}$ (cf. \ref{eq:sdissb}).

Furthermore, the \emph{bare} dimensionless  couplings are:
\begin{align}
\tilde{g}_D(0) &=   g^2_{BF } D(q=0), \\
g_{b D}(0) &= 2 g^2_{BF} \mathcal{B}^2_1 \rho^2 a_0 D(q = 2k^B_F),\\
g_{u D}(0) &= 2 g^{2}_{BF}\mathcal{B}^2_1 \rho^2 a_0 D(q = 2 k^B_F).
\end{align}
In the above expressions we have made explicit the dependence of the couplings of the cut-off scale, $a_0$
through the parameter $\ell = \log \frac{a_0(\ell)}{a_0}$, that is $a_0(\ell) = e^{\ell} a_0$ and thus $\ell = 0$
corresponds to the scale of the bare cut-off $a_0 \approx v \tau_c$, being $\tau_c$ the short-time cut-off
introduced earlier.

\section{Renormalization Group  Analysis}\label{sec:rg}

Physically, the renormalization group (RG) flow of a system describes its
behavior as it is cooled down towards the absolute zero. The effect of temperature
can be mimicked by decreasing the short wavelength cutoff $\sim \frac{\hbar}{a_0}$
introduced to properly define the low-temperature effective model of the last section.
As the absolute temperature decreases, the ground state is approached, and the
couplings that define the effective low-energy theory of equations~\eqref{eq:sb0}, \eqref{eq:su}, \eqref{eq:sdissf}
\eqref{eq:sdissb}, etc. (i.e., $K$, $v$, $g_u$, $g_{bD}$, ...),
must change accordingly in order to account for the reduction of the available excited states.
Thus, the quantum phases of the system can be studied by analyzing the asymptotic behavior of the
`flow' of these couplings in the limit where the cut-off tends to zero, that is, as the
absolute temperature vanishes.  In the perturbative approach to  RG,
the flow is described by a set of differential equations, whose
solutions we study in this section.

Simple power-counting arguments show that
$S^f_D \sim \int dq d\omega \, q^2 |\omega| \: |\phi(q,\omega)|^2$ is an
irrelevant perturbation in the  renormalization-group sense. This is true provided $D(q=0)$ is not singular,
which is indeed the case (see Fig.~\ref{fig:imchi} and Appendix~\ref{app:response}).
Indeed, this term alone leads to a momentum dependent broadening of the long-wave
length phonon excitations of the gapless phase of the model in Eq.~\ref{eq:sb0}). Therefore,
in order to study the low-temperature properties of the model, it is justified to drop $S^f_{D}$,
and therefore  we shall next focus our attention on the second term in Eq.~\eqref{eq:sbd} and
consider the effective model described by $S = S_{B} + S^b_{D}$, where $S_{B}$ is given
by Eq.~\eqref{eq:sb0} and $S^{b}_D$ given by Eq.~\eqref{eq:sdissb}. In the half-filled
case, we also have to take into account $S^u_D$ given by Eq.~\eqref{eq:sdissu}.
The resulting action contains only marginal and (potentially) relevant  perturbations in
the RG  sense,  which we shall analyze in this section.
In what follows, we shall consider the cases of integer and half-integer lattice filling separately.
The  details of the perturbative derivation of the RG equations are given in the Appendix~\ref{app:rg}.

\subsection{Integer Lattice filling}\label{sec:intf}

 To $O(g_{bD}, g^2_u)$ the  flow equations in this case read:
\begin{align}
\frac{d g_u}{d\ell} &= (2 - K) g_u,\\
\frac{dg_{bD}}{d\ell} &= (1 - 2 K ) g_{bD}, \\
\frac{d K}{d\ell} &=  -  (g^2_u+ 2 \pi g_{bD}) K^2,\\
\frac{d v}{d\ell} &=  -2 \pi g_{bD} K v.
\end{align}
We neglect terms of $O(g^2_{bD})$ or higher because $g_{bf D}(0) \propto g^2_{BF}$, that is,
$g_{b D}$ is already second order in the Bose-Fermi coupling, which is  assumed to be small.
For $g_{bD} = 0$, the equations reduce to those of a pure 1D Boson system in a commensurate potential
first obtained by Haldane~\cite{Haldane} (see also~\cite{review1D,giambook});
for $g_u = 0$, the equations reduce to those derived in Ref.~\onlinecite{Cazalilla06}, which describe
the quantum phase transition between a Tomonaga-Luttinger liquid  and a dissipative insulator (DI).

 The above equations show that near the SF to MI quantum critical point (corresponding to $K^*  =  2$, $g_u = 0$, $g_{bD} =0$)
 the dissipative interaction is a highly irrelevant operator because $1- 2 K \approx -3$. Thus,
 the most important effect of the Fermi component of the mixture is to introduce a renormalization
 of the periodic potential and the screening of the interactions, which leads to the
 renormalization of the Luttinger parameter $K$ and the sound velocity $v$ given
 by Eq.~\eqref{eq:renparam}.

  From the analysis of the RG equations, which implies that the dissipation is an irrelevant operator in the RG sense,
 we conclude that dissipative effects are weak in the MI phase where $g_u$ grows as the energy cut-off $\frac{\hbar v e^{-\ell}}{a_0 }$
 ($\sim$ the absolute temperature) decreases. Thus, the dissipative term can be treated using perturbation theory,
 and leads to a small (when compared to the excitation energy) broadening of the phonon excitations in the superfluid TLL phase.
As for the excitations of  the MI phase, which  corresponds to a `particle' (i.e.~excess by one bosons) or a `hole' (i.e. absence of bosons)
 propagating against the Mott-insulating background, the dissipative part of the interaction with the Fermi gas similarly
introduces  damping on their motion, which translates into the
broadening of the excitation energy dispersion.  Such enhancement of the excitation
broadening can be measured by lattice modulation
spectroscopy~\cite{oplatt1,pinning1d_exp,Kollath2006,Iucci06}.

\subsection{Half-Integer Lattice filling}\label{sec:halfint}
\begin{figure}[t]
\includegraphics[height=0.3\textwidth]{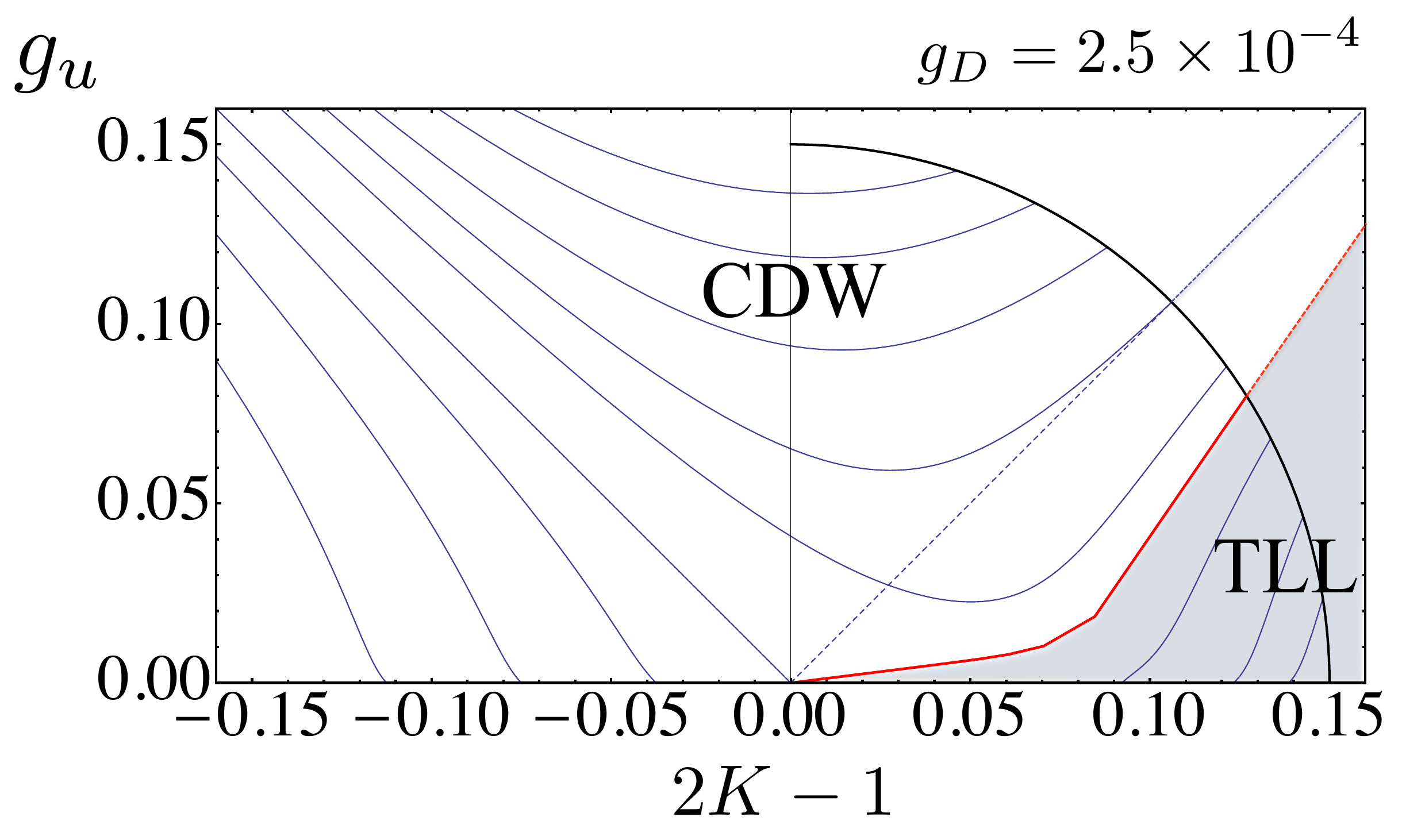}
\caption{\label{fig:phasediag} Phase diagram for the Tomonaga-Luttinger liquid (TLL) to Charge Density Wave (CDW) transition
in the presence of a Fermi gas for $g_D \simeq 2.5 \times 10^{-4}$, which corresponds to $|g_{BF}|/\mu_B \sim 10^{-2}$ ($\mu_B$ being the chemical potential of the bosons). $K$ is the Luttinger
parameter of the bosons in the mixture (cf. Eq.~\ref{eq:renparam})
and $g_u \propto U_{B\parallel} + O(g_{BF})$, where $U_{B\parallel}$ is the external periodic potential. The
shaded area is the TLL phase.  The diagonal dashed line represents the TLL-CDW phase boundary in the absence of fermions.
The curves in the diagram represent RG flows for the $K$ and $g_u$ couplings for a set of initial conditions lying on the quarter circle
on the right. The flow proceeds from right to left as as $K$ always decreases according to Eq.~\ref{eq:rg13}.}
\end{figure}

\begin{figure}[t]
\includegraphics[height=0.3\textwidth]{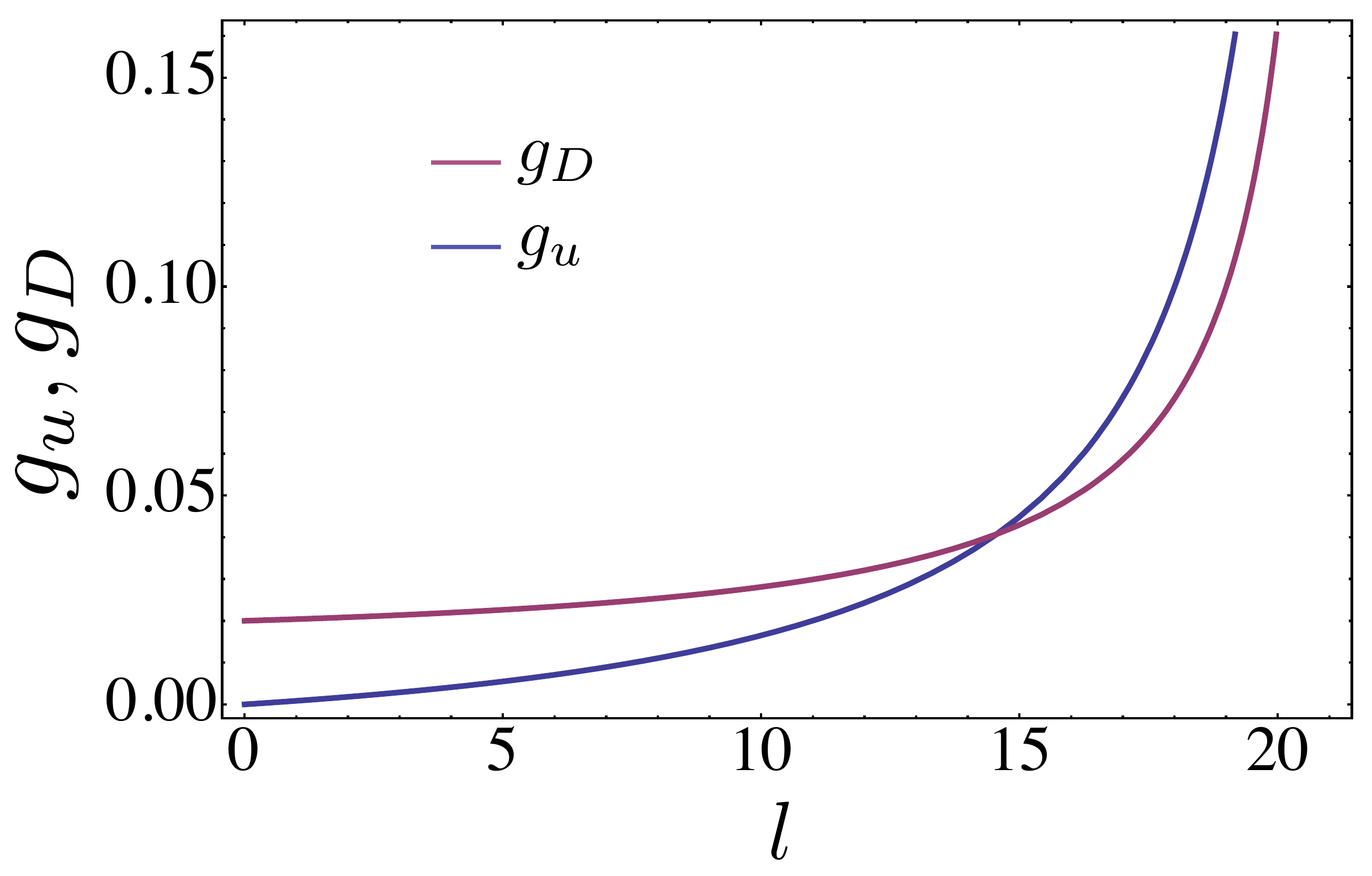}
\caption{\label{fig:gugd} Runaway renormalization-group (RG) flow of the couplings $g_{u}$, representing the periodic potential,
and $g_{D}$, representing the effect of the fermion-induced dissipation for $K \lesssim \frac{1}{2}$. 
We found that even for relatively small initial
potential $g_u(0)$ the RG flow of $g_u(\ell)$ eventually overcomes the flow of $g_D(\ell)$ and becomes $g_u(\ell) \sim 1$ first.
This means that the system localizes and becomes a Mott insulator.
However, as this plot illustrates, the effect of $g_D$ , i.e. the renormalized dissipative coupling, is not negligible.}
\end{figure}

In this case, and given that the initial conditions are the same for the $S^b_{D}$ and $S^b_{D}$ we
note that they can be combined into a single term $S_{D}[\phi] = S^b_D[\phi] + S^u_{D}[\phi] $, which can be written as:
\begin{equation}
S_{D}[\phi] = \frac{g_D}{2 a_0} \int dx d\tau d\tau^{\prime}
\frac{\left[ \cos 2 \phi(x,\tau) - \cos 2\phi(x,\tau^{\prime}) \right]^2}{\left( |\tau-\tau^{\prime}| + \tau_c \right)^2},\label{eq:sdissd}
\end{equation}
where $g_D(0) = \frac{1}{2} \left[ g_{b D}(0) +  g_{u D}(0) \right]$. The RG flow equations for this system
then read:
\begin{align}
\frac{d g_u}{d\ell} &= (2 - 4 K) g_u + \pi g_D,\label{eq:rg11}\\
\frac{dg_{D}}{d\ell} &= (1 - 2 K + 4 g_u ) g_{D},\label{eq:rg12} \\
\frac{d K}{d\ell} &=  -  (4 g^2_u+ 2\pi g_{bD}) K^2,\label{eq:rg13}\\
\frac{d v}{d\ell} &=  -2 \pi g_{D} K v.\label{eq:rg14}
\end{align}
These RG equations describe the flow in the vicinity of a quantum
critical point located at $K^* = \frac{1}{2}$, $g^*_u = g^*_D = 0$.
Integrating them numerically, we obtain the phase diagram depicted
in Fig.~\ref{fig:phasediag}.  Thus, we find that, for a relatively
weak boson-fermion coupling $|g_{BF}|/\mu_B \sim 10^{-2}$, the part
of the phase diagram occupied by the SF Tomonaga-Luttinger liquid
phase (TLL) shrinks considerably.  The latter phase is identified by
the RG flows for which both $g_u$ and $g_D\to 0$ as the phonon
cut-off ${\hbar}{a_0 e^{\ell}}$ is reduced to zero (i.e.~for $\ell
\to +\infty$), that is, as the absolute temperature is decreased. On
the other hand, the CDW phase is identified with those flows for
which $g_u \sim 1$ at a certain value of $\ell^*$.
 However, it is also worth noticing that we have observed numerically (see Fig.~\ref{fig:gugd})
  that, especially close to the phase boundary
 (red curve in Fig.~\ref{fig:phasediag}), $g_u(\ell^*)/g_D(\ell^*)\sim 1$, even if $g_u$ becomes of order one
 first in all cases studied. This means that, even if the low-energy physics of this phase is dominated by the
 potential term $\propto g_u$, the dissipative effects are by no means negligible.  It is interesting that this happens
 independently of how small the bare $g_u(0)$ is, and even in the limit $g_u(0) \to 0^+$. This is because,
ultimately, the RG flow of $g_u(\ell)$ is controlled by the first term in Eq.~\eqref{eq:rg11}, which leads to a much
faster growth, although for  small $g_u(0)$, the initial flow may be controlled by the second term in Eq.~\eqref{eq:rg11}.

\begin{figure}[t]
\includegraphics[height=0.3\textwidth]{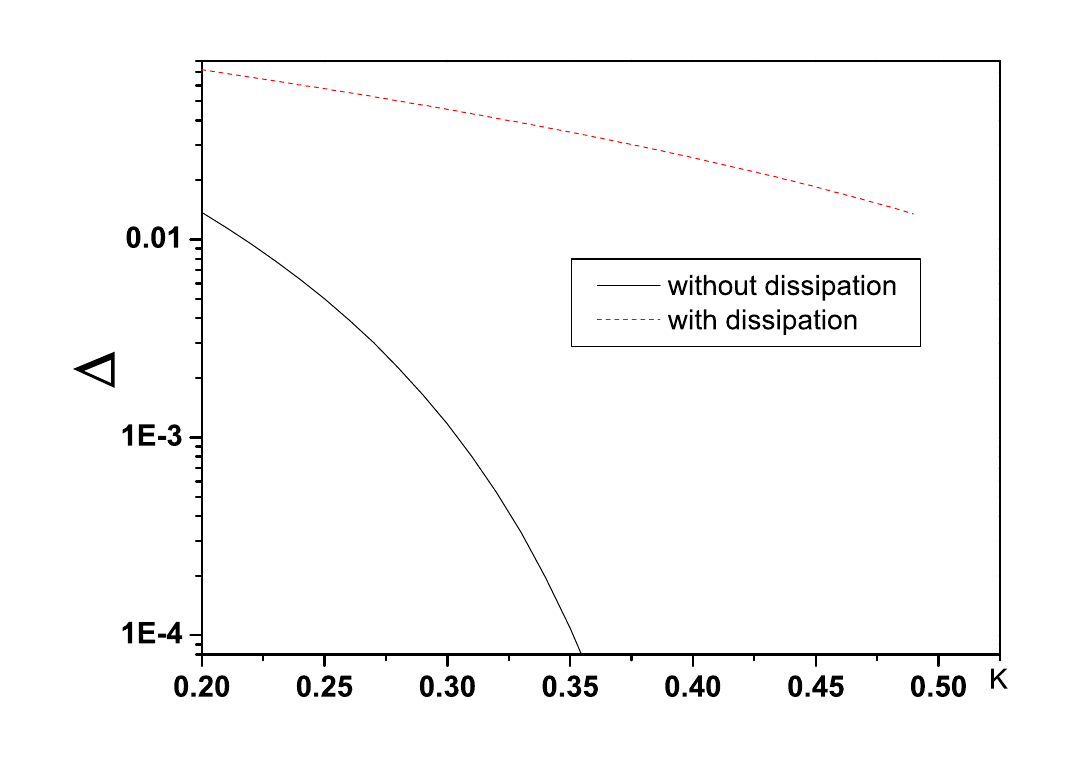}
\caption{\label{fig:gap}  Mott gap $\Delta$ in the presence and absence of fermion-induced dissipation
as obtained from the self-consistent harmonic approximation (SCHA, see Sec.~\ref{sec:halfint} for details). It can be seen
that the dissipation greatly enhances the Mott gap by suppressing the quantum fluctuations of the bosons in the CDW Mott insulating
state. Note that the SCHA erroneous yields a discontinuous  phase transition at the critical point $K^{*} = \frac{1}{2}$. This is a well known artifact of this approximation~\cite{giambook}.}
\end{figure}

The RG flow equations indicate that the quantum phase transition occurs at K = 1/2, where the dissipation and periodic potential simultaneously become relevant, and the system is driven from a superfluid to a CDW Mott-insulating states. To study the interplay between the dissipation and interaction around the critical point, we adopt a variational self-consistent harmonic approximation (SCHA) by choosing a trial effective action of the from:
\begin{align}
&S_{v}[\phi]=\int \frac{dq d\omega}{(2\pi)^2}G_v^{-1}(q,\omega)\phi^*(q,\omega)\phi(q,\omega)
\label{eq:strial}
\end{align}
where we have defined the Green's function $G_v(q,\omega)=\Big[\frac{1}{2\pi K} (\frac{\omega^2}{v_s} + v_sq^2)\Big] + \frac{\eta}{a_0}|\omega| +  \frac{\Delta}{a_0\tau_c}\Big]^{-1}$ with the dimensionless self-consistent parameters $\eta$ and $\Delta$  that can be determined by the minimization of the variational free-energy.  A variational estimate $F_{var}$ of the true free-energy $F$ can be
obtained from Feynman's variational principle~\cite{giambook}:
\begin{align}
F\leq F_{var}=F_v + \beta^{-1} \langle S-S_v\rangle_v
\label{eq:fvar}
\end{align}
Therefore, optimizing $\delta F_{var}[G_v]/\delta G_v=0$,) the parameters $\eta$ and $\Delta$ are found by solving the self-consistent equation above (Eq.~(\ref{eq:self})), so that (see appendix~\ref{app:scha} for further details):
\begin{align}
&\eta = \frac{8g_{u}}{(2\pi)^2}\alpha^2(\eta,\Delta,K),\label{eq:scha1}\\
&\Delta=\frac{8(g_{u}+g_D)}{(2\pi)^2}\alpha^2(\eta,\Delta,K),\label{eq:scha2}
\end{align}
where we have introduced $\alpha(\eta,\Delta)=\Big[\frac{\eta K\pi + 2\sqrt{K\pi\Delta}}{4}\Big]^{2K}$. The numerical solution
of these equation for the gap $\Delta$ is shown in Fig.~\ref{fig:gap}. It can be seen that  the gap is 
enhanced for $K < \frac{1}{2}$. This expected is because quantum dissipation is akin to classical friction, which hinders the motion of the particles and thus helps to stabilize the CDW Mott-insulating state. Note, however, that the SCHA erroneously yields a discontinuous transition for $K = \frac{1}{2}$. This is a well known artifact of this approximation~\cite{giambook}.

\section{Commensurate - Incommensurate transition in the Presence of Dissipation}\label{sec:cic}
\subsection{Integer filling}\label{sec:integ}

 In this case, as for the TLL to MI transition, the effect of dissipation is rather weak. A way of understanding this
is to stop the RG flow when $g_u(\ell) \sim 1$ and consider the sine-Gordon model at the Luther-Emery point where
it maps to a 1D relativistic model of massive  (Dirac) fermions~\cite{Gogolin,giambook}. Diagonalization of this model yields two
bands separated by a gap: a filled `valence' band and an empty `conduction' band~\cite{Gogolin,giambook}.
  Tuning  the chemical for the bosons
amounts to introducing particles in the conduction band or holes in
the valence band~\cite{Gogolin}. For small particle (hole) density,
the system can be described as a Tonks-Girardeau
gas~\cite{Girardeau65}  characterized by Luttinger parameter $K
\simeq 1$.  The dissipation being an irrelevant for $K >  K^* =
\frac{1}{2}$, its effect on such a dilute liquid of particles
(holes) is negligible as far as the ground state properties are
concerned (although it will lead to a small linewidth of the
excitations, which is due to collisions between the bosons and the
fermions). Thus, in particular,  the exponents characterizing the
commensurate to incommensurate (C-IC) transition are thus expected
to remain unchanged and, therefore, the density of particles (or
holes)~\cite{Gogolin,review1D} will grow as $\sqrt{\mu-\mu_c}$,
where $\mu_c \sim \Delta$, where $\Delta$ is the MI gap.

\subsection{Half-integer filling}\label{sec:halfint2}
 For half-integer filling the situation is very different, as it was already pointed out in our discussion of the previous
section. We can realize this by considering again the case where we take $g_D$ infinitesimally small but $g_u\sim 1$.
Applying the same reasoning used in the previous section, the sine-Gordon model $S_{b}[\phi] + S_{u}[\phi]$ in this case maps to
a system of Dirac fermions describing the (fractionally charged) soliton and  anti-soliton excitations of the CDW state
(configurations of the form $10101011010101$, for the solitons, and $101010010101$, for the anti-solitons).
A dilute gas of such excitations  can be described as a Luttinger gas with a parameter $K \simeq \frac{1}{4}  < K^* = \frac{1}{2}$.
Thus, the dissipative term $S_{D}[\phi]$ from Eq.~\eqref{eq:sdissd} is a strongly relevant perturbation, which,
as discussed in Ref.~\cite{Cazalilla06}, leads to the localization of the system in a new phase, which we term
dissipative insulator (DI). In this phase, the boson density, $\langle\rho_B(x) \rangle$ exhibits 
long-range order~\cite{Cazalilla06} with a characteristic wave number equal to $4\pi \rho_0$.

 However, it is worth mentioning that, as Fig.~\ref{fig:gugd}  demonstrates, assuming that $g_D$ is infinitesimal when
 $g_D \sim 1$ is not representative of the  the RG flow described in the previous section. Indeed, we  numerically found that even in the case $g_u(0) \to 0$, $g_D(\ell^*) \lesssim g_u(\ell^*) \sim 1$ (see Fig.~\ref{fig:gugd}) in other words, the dissipation,
 although diverging less strongly than the periodic potential, is not a small perturbation on the CDW state. Thus, we expect
 that the
 dissipative term needs to be treated on equal footing with the
potential term $\propto g_u$. The universality class of the
 commensurate to incommensurate transition  is therefore expected to be different from the case of integer filling.

\section{Conclusions}\label{sec:conclusions}

 In conclusion, we have studied a model for a mixed dimensional Bose-Fermi mixture in an optical lattice,
 where the bosons are confined to one dimension whereas the fermions are free to hop in three dimensions
 (albeit with renormalized dispersion). We have argued
that this system is a realization of a 1D interacting Bose gas coupled to a dissipative bath of the Ohmic type. In addition, the fermions
also screen the boson-boson interactions. For integer  filling of the boson lattice, 
 we have found that the dominant effect of the fermions on the bosons is the screening of their
interactions,  as it was also observed in mean-field studies of 3D dimensional optical lattices~\cite{screendiss2}.
Thus, provided the so-called self-trapping effect can be subtracted or compensated, the screening of the boson interactions leads
to an enhancement  of the superfluid properties as the bosons become polarons with reduced effective interactions.
In this case, dissipation effects only contribute
to an increase in the linewidth of the excitations in both the superfluid and Mott-insulating phases, which  could be detected
by means of lattice modulation spectroscopy~\cite{oplatt1,pinning1d_exp,Iucci06}.
\begin{figure}[b]
\includegraphics[height=0.40\textwidth]{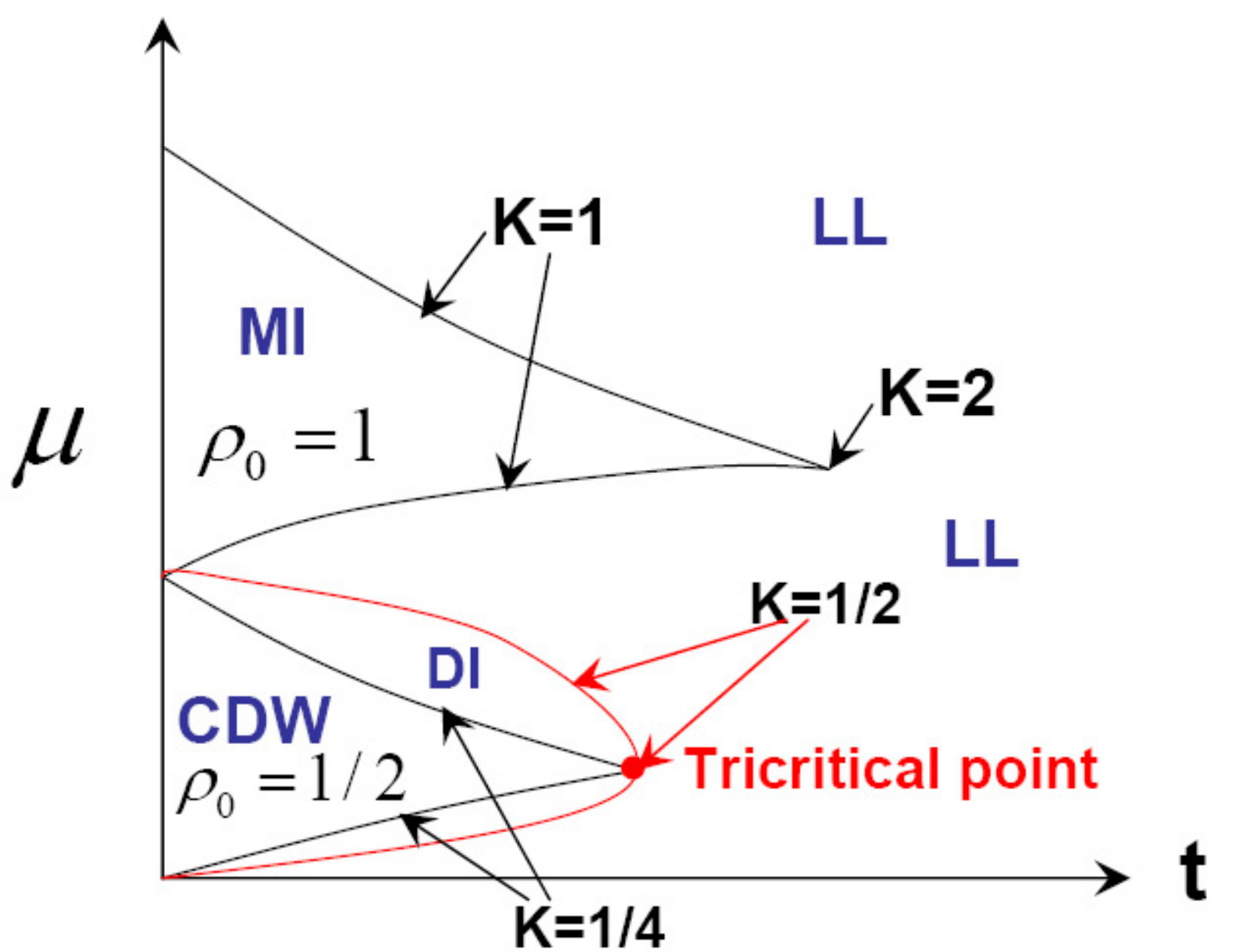}
\caption{\label{fig:schem} Schematic phase diagram of a  system of heavy bosons confined to 1D and 
and coupled to a dissipative bath of light fermions moving in 3D.}
\end{figure}

 On the other hand, the effect of the fermion-induced dissipation is much more severe when the bosons are close to a
superfluid to CDW Mott-insulator transition, which happens at half-integer filling. In this case, the dissipative effects
strongly hinder the motion of the bosons and help stabilizing the CDW phase (cf. Fig.~\ref{fig:phasediag}) as well as
enhancing the CDW gap (cf. Fig.~\ref{fig:gap}). This effect  leads to a dramatic suppression of the superfluid phase relatative
to the pure boson case, which can observed as a reduction of the potential depth required for the bosons to
localize in the CDW phase. The enhancement of the gap on the CDW side of the transition can be  also
probed using lattice modulation spectroscopy.

 We have also studied the commensurate-incommensurate transition and  argued that in the case of integer
 lattice filling, the fermion-induced dissipation is an irrelevant perturbation and therefore, the universality class should
 not be altered. However, in the case of half-integer filling the dissipation is relevant (but less than the external  potential)
 and therefore we expect the universality class will be modified. This subject requires further study, but it will not
 be pursued here. The conclusions of this work  are summarized in the schematic phase 
 diagram of Fig.~\ref{fig:schem}.  

\section{Acknowledgement }\label{sec:Acknowledgement}

 EM acknowledges support from the CSIC JAE-predoc program, co-financed by
the European Science Foundation. EM and MAC also acknowledge the
support of the Basque Departamento de Educaci\'on, the UPV/EHU (Grant
No.  IT-366-07), and the Spanish MINECO (Grant No. FIS2010-19609-CO2-02).
US and ZC acknowledge support from DFG through FOR801.

\appendix

\section{Relating $\chi^R_F(x,\omega)$ to $\chi_F(x,\tau)$} 
\label{app:relating}

In this appendix we will derive the identity that we used in the main text to relate the retarded
density correlation function to its imaginary time version at zero temperature. We shall first
recall that the retarded correlation function is defined as:
\begin{equation}
\chi^R_F(x,t) = -\frac{i}{\hbar} \vartheta(t)\: \langle  \left[ \delta \rho_F(x,t) \delta \rho_F(0,0)\right] \rangle_F,\label{eq:ret}
\end{equation}
where $\delta \rho_F(x,t) = \int d\mathbf{r}_{\perp} \, |w_0(\mathbf{r}_{\perp})|^2 \, \delta \rho_F(x,\mathbf{r}_{\perp},t)$,
$\delta\rho_F(x,t) = e^{iH_F t/\hbar} \delta \rho_F(x) e^{-iH_F t/\hbar}$,
and $\delta \rho_F(\mathbf{r}) = \rho_F(\mathbf{r}) - \rho^0_F(\mathbf{r})$,. However, the imaginary
time correlation is defined as:
\begin{equation}
\chi_F(x,\tau)= -\frac{1}{\hbar} \langle \delta \rho_F(x,\tau) \delta \rho_F(0,0)\rangle_F,
\end{equation}
where $\delta\rho_F(x,\tau) = e^{H_F \tau/\hbar} \delta \rho_F(x) e^{-H_F \tau/\hbar}$. By taking the
Fourier transform of the spectral representation of  \eqref{eq:ret} and comparing it to the spectral
representation of
\begin{equation}
\chi_F(x,i\omega_n)  = \int^{\hbar\beta/2}_{-\hbar\beta/2} d\tau \, \chi_F(x,\tau)\: e^{i\omega_n \tau},
\label{eq:aux}
\end{equation}
we arrive at the following relation:
\begin{align}
\chi_F(x,i\omega_n) &= \int \frac{d\omega}{\pi} \: \frac{\mathrm{Im}\: \chi^R_F(x,\omega)}{\omega - i\omega_n} \\
 &=  \int^{+\infty}_0 \frac{d\omega}{\pi}\: \left[ \frac{\mathrm{Im}\: \chi^R_F(x,\omega)}{\omega - i \omega_n}  + \frac{ \mathrm{Im}
\: \chi^R_F(x,\omega)}{\omega + i\omega_n} \right],\quad
\end{align}
where, in the deriving the last expression, we have used that $\mathrm{Im} \: \chi_F(x,-\omega) = - \mathrm{Im} \: \chi_F(x,\omega)$.
Hence, introducing the last expression in   \eqref{eq:aux}, taking $\beta \to +\infty$, and performing the integral
over $\omega_n$ with the help of Jordan's lemma, we arrive at the desired
result:
\begin{equation}
\chi_F(x,\tau) = \int^{+\infty}_0 \frac{d\omega}{\pi} \, e^{-\omega |\tau|} \: \mathrm{Im} \: \chi^R_F(x,\omega).
\end{equation}
\section{Fermion bath response function}\label{app:response}

 Let us  consider  the Fourier transform of the density response of the Fermi gas at
 zero temperature,   which, as we neglect the interactions induced by the
bosons on the fermions, is just the Lindhard function.
 Recalling that the Matsubara version of the latter is defined as $\chi_F(x, \mathbf{r}_{\perp}, \mathbf{r}^{\prime}_{\perp},\tau)
= - \frac{1}{\hbar} \langle \delta \rho_F(x,\mathbf{r}_{\perp},\tau)\delta \rho_F(0,\mathbf{r}^{\prime}_{\perp},0) \rangle_F$,
where $\delta \rho_F(x,\mathbf{r}_{\perp})  = \rho_F(x,\mathbf{r}_{\perp}) - \rho^0(x,\mathbf{r}_{\perp})$, being
 $\rho_F(x,\mathbf{r}_{\perp}) = \sum_{k,k^{\prime},\mathbf{k}_{\perp},\mathbf{k}^{\prime}_{\perp}}
\varphi^*_{k,\mathbf{k}_{\perp}}(x,\mathbf{r}_{\perp})  \varphi_{k^{\prime},\mathbf{k}^{\prime}_{\perp}}(x,\mathbf{r}_{\perp}) \:
f^{\dag}_{k,\mathbf{k}_{\perp}} f_{k^{\prime},\mathbf{k}^{\prime}_{\perp}}$ the density operator and
$\rho^0(x,\mathbf{r}_{\perp})  = \langle \rho_F(x,\mathbf{r}_{\perp}) \rangle_F$, the equilibrium density.
We shall assume that  the single  particle orbitals of the fermions are given by
\begin{align}
\varphi_{k,\mathbf{k}_{\perp}}(x,\mathbf{r}_{\perp})  &= \varphi_{k}(x) \varphi_{\mathbf{k}_{\perp}}(\mathbf{r}_{\perp}) \notag\\
&= \frac{1}{\sqrt{L M}} \sum_{\mathbf{R}} e^{i (k x  + \mathbf{k}_{\perp} \cdot \mathbf{R})}\,
w^F_0(\mathbf{r}_{\perp} - \mathbf{R}), \label{eq:fermiorbital}
\end{align}
where $L$ is the (normalization) length in 1D
and $M$ is the number of  lattice sites  labelled by $\mathbf{R} = (n,m) b_0$ ($b_0$ is the
lattice parameter), and  $w^F_0(\mathbf{r}_{\perp})$ is Wannier orbital
for the fermions. In the above expression we have assumed  that the strength of the
longitudinal potential in 1D is weak so that the Bloch orbitals $\varphi_k(x) \simeq \frac{e^{ik x}}{\sqrt{L}}$.
Thus, we arrive at the following expression:
\begin{align}
\chi_F(q,\mathbf{r}_{\perp}, \mathbf{r}^{\prime}_{\perp},\omega) &=  \int d\tau\, e^{i(\omega \tau - q x)} \,
\chi_F(x, \mathbf{r}_{\perp}, \mathbf{r}^{\prime}_{\perp},\tau)\notag \\
&=  \sum_{k,\mathbf{k}_{\perp},\mathbf{k'}_{\perp}}
\frac{n_{k,\mathbf{k}_{\perp}} - n_{k+q,\mathbf{k'}_{\perp}}}{i\hbar \omega - \epsilon(k+q,\mathbf{k}'_{\perp})  + \epsilon(k,\mathbf{k}_{\perp}) } \, \notag\\
&\times A_{\mathbf{k}_{\perp},\mathbf{k'}_{\perp} }(\mathbf{r}_{\perp}, \mathbf{r}^{\prime}_{\perp}),\label{eq:lindhard}
\end{align}
where the function $A_{\mathbf{k}_{\perp},\mathbf{k'}_{\perp} }(\mathbf{r}_{\perp}, \mathbf{r'}_{\perp}) = \varphi^*_{\mathbf{k}_{\perp}}(\mathbf{r}_{\perp}) \varphi_{\mathbf{k}_{\perp}}(\mathbf{r'}_{\perp})\varphi_{\mathbf{k'}_{\perp}}(\mathbf{r}_{\perp})\varphi^*_{\mathbf{k'}_{\perp}}(\mathbf{r'}_{\perp})$.   The single-particle dispersion of the fermions is
\begin{align}
\epsilon(k, \mathbf{k}_{\perp}) &= \epsilon_{\|}(k) + \epsilon(\mathbf{k}_{\perp})  \notag\\
 &=  \frac{\hbar^2 k^2}{2m^*_F} - 2 t_{\perp} \left( \cos k_y b_0
+ \cos k_z b_0  \right)
\end{align}
where we have assumed that the longitudinal  dispersion is approximated by a quadratic dispersion
characterized by an effective mass $m^*_F\approx m_F $  and transverse motion is described by a tight-biding  dispersion
characterized by a transverse hopping $t_{\perp}$.

 Indeed, the response function in which we are interested is not the Lindhard function, \ref{eq:lindhard}, but
the following integral of it:
\begin{equation}
\chi_F(q,\omega_n) = \int d\mathbf{r}_{\perp} d\mathbf{r}^{\prime}_{\perp}\,
F_0(\mathbf{r}_{\perp},\mathbf{r}^{\prime}_{\perp}) \chi(q,\mathbf{r}_{\perp},
\mathbf{r}^{\prime}_{\perp},\omega_n),\label{eq:resp2}
\end{equation}
where $F_0(\mathbf{r}_{\perp},\mathbf{r}^{\prime}_{\perp}) = |w_0(\mathbf{r}_{\perp}) w_0(\mathbf{r}_{\perp})|^2$, where
$w_0(\mathbf{r})$ are the Wannier orbitals for the bosons in the lowest Bloch band. Thus, in order to compute \eqref{eq:resp2},
we need to consider the following integral:
\begin{align}
&\int d\mathbf{r}_{\perp} d\mathbf{r'}_{\perp}   \: F_0(\mathbf{r}_{\perp},\mathbf{r}^{\prime}_{\perp}) \: A_{\mathbf{k}_{\perp},\mathbf{k'}_{\perp} }(\mathbf{r}_{\perp},\mathbf{r'}_{\perp}) \notag \\
& =   \left| \int d\mathbf{r}_{\perp} \varphi^*_{\mathbf{k}_{\perp}}(\mathbf{r}_{\perp}) |w_0(\mathbf{r}_{\perp})|^2 \varphi_{\mathbf{k'}_{\perp}}(\mathbf{r}_{\perp})  \right|^2  \\
&= \Big| \frac{1}{M} \sum_{\mathbf{R},\mathbf{R'}} e^{i \left( \mathbf{k}_{\perp}\cdot \mathbf{R} - \mathbf{k'}_{\perp} \cdot \mathbf{R'} \right)}
\int d\mathbf{r}_{\perp}\, |w_0(\mathbf{r}_{\perp})|^2 \notag \\
&\times    \left[ w^F_0(\mathbf{r}_{\perp} - \mathbf{R})\right]^*
 w^F_0(\mathbf{r}_{\perp} - \mathbf{R'})   \Big|^2 \\
 &\simeq  \left| \frac{1}{M}   \int d\mathbf{r}_{\perp}\, |w_0(\mathbf{r}_{\perp})|^2 \left| w^F_0(\mathbf{r}_{\perp})\right|^2
   \right|^2 = \frac{A}{M^2},
  \end{align}
where we have approximated $w_0(\mathbf{r}_{\perp}) \simeq e^{-|\mathbf{r}_{\perp}|^2/2\ell^2_{B\perp}}/(2\pi \ell^2_{B\perp})$
 and $w^F_0(\mathbf{r}_{\perp}) \simeq e^{-|\mathbf{r}_{\perp}|^2/2\ell^2_{F\perp}}/(2\pi \ell^2_{F\perp})$ and assumed
 that $\ell_{B\perp} \ll \ell_{F\perp}$, so that we can neglect overlap between the  Wannier orbitals
 for $\mathbf{R}\neq \mathbf{R}^{\prime}$.  In the above expression,
\begin{equation}
A = \int d\mathbf{r}_{\perp}\, |w_0(\mathbf{r}_{\perp})|^2 \left| w^F_0(\mathbf{r}_{\perp})\right|^2 =  \frac{1}{\pi^2 (\ell^2_{F\perp}
+ \ell^2_{B \perp})^2}.\quad
\end{equation}
 Hence,
\begin{align}
\chi_F(q, i \omega_n) &\simeq
\frac{A}{M^2 L } \sum_{k,\mathbf{k}_{\perp},\mathbf{k'}_{\perp}}
\frac{n_{k,\mathbf{k}_{\perp}} - n_{k+q,\mathbf{k'}_{\perp}}}{i\hbar \omega_n - \epsilon(k+q,\mathbf{k}'_{\perp})  + \epsilon(k,\mathbf{k}_{\perp}) } .
\end{align}

 Next, we  take the thermodynamic limit,  transform the sums over $k,\mathbf{k}_{\perp},\mathbf{k'}_{\perp}$
into integrals, and introduce the density of states of the 2D (square) lattice of tubes~\cite{Economou},
\begin{equation}
\rho(\varepsilon) = \frac{2}{\pi^2 W} K\left[\sqrt{1 - \left(\frac{\varepsilon}{W}\right)^2}\right]\: \theta\left(W^2 - \epsilon^2\right).
\end{equation}
where $K(z)$ denotes  the complete elliptic integral of the first kind and $W  = 4 t_{\perp}$. Thus,
the retarded response function (obtained from $\chi_F(q,i\omega_n)$ by means of analytic continuation where
 $i\omega_n \to \omega^{+} = \omega  + i 0^{+}$) can be rewritten as follows
\begin{align}
\chi^{R}_F(q,  \omega_n)  &= A  \int^{+W}_{-W} d\varepsilon d\varepsilon'\,  \rho(\varepsilon) \rho(\varepsilon') \notag\\
&\times  \int \frac{dk }{2\pi} \frac{n_{k,\varepsilon} - n_{k+q,\varepsilon'}}{\hbar \omega^{+} + \varepsilon-\varepsilon' - \epsilon_{\parallel}(k+q) + \epsilon_{\parallel}(k)} \\
& =  A  \int^{+W}_{-W} d\varepsilon d\varepsilon'\,  \rho(\varepsilon) \rho(\varepsilon') \notag\\
& \times \int \frac{dk }{2\pi}  n_{k,\varepsilon}
\left[ \frac{1}{\hbar\omega^{+}  + \varepsilon-\varepsilon' - \epsilon_{\parallel}(k+q) + \epsilon_{\parallel}(k)} \right.  \notag\\
& \quad \left.  + \frac{1}{-\hbar\omega^{+}  + \varepsilon-\varepsilon' +\epsilon_{\parallel}(k+q)  - \epsilon_{\parallel}(k)} \right] \nonumber\\
& =  A  \int^{+W}_{-W} d\varepsilon d\varepsilon'\,  \rho(\varepsilon) \rho(\varepsilon') \, \int \frac{dk }{2\pi}  n_{k,\varepsilon} \notag\\
& \times  \left[ \frac{1}{\hbar\omega^{+}  + \varepsilon-\varepsilon' - \epsilon_{\parallel}(k+q) + \epsilon_{\parallel}(k)} + \right. \notag\\
&\quad \left.  \left( \omega^{+}\to -\omega^{+}\right) \right]
\end{align}
At zero temperature $n_{k,\varepsilon} = \theta(\epsilon_F - \epsilon - \epsilon_{\parallel}(k))$, where $\epsilon_F  = \mu_F(T= 0)$ is the Fermi energy (note that $\epsilon_F >  -W$ otherwise there will be no fermions in the mixture).

 Let us first consider (minus) the imaginary part of $\chi^R_F(q,\omega)$:
\begin{align}
\mathrm{Im} \left[ - \chi^R_F(q,\omega) \right]
&=   \frac{A}{2}\int^{+W}_{-W} d\varepsilon  \, \int dk \,
\,  \theta\left(\epsilon_F - \epsilon - \frac{\hbar^2 k^2}{2m^*_F}\right)  \notag\\
& \times \rho(\varepsilon) \left[ \rho\left(\hbar \omega +\varepsilon  - \frac{\hbar^2 q^2}{2m^*_F} - \frac{\hbar^2 k q}{m^*_F} \right)  \right. \nonumber\\
 &\quad\quad  \left.- \rho\left(\hbar \omega -\varepsilon + \frac{\hbar^2 q^2}{2m^*_F} + \frac{\hbar^2 k q}{m^*_F}\right)
 \right],  \quad\quad
\ \end{align}
where we have set $\epsilon_{\parallel}(k+q) - \epsilon_{\parallel}(k) = \frac{\hbar^2 q^2}{2m^*_F} + \frac{\hbar^2 k q}{m^*_F}$
The above expression can be used to obtain the (imaginary part of the) response for arbitrary $\omega$. However, we are only
interested in the regime of small $\omega$, for which we can expand $\rho\left(\hbar \omega \pm E(k,q,\varepsilon) \right)=
\rho(E(k,q,\varepsilon)) \pm \rho^{\prime}(E(k,q,\varepsilon)) \hbar\omega + \cdots$ (where $E(k,q,\varepsilon)
= \varepsilon - \frac{\hbar^2}{2m^*_F}(q^2 + 2k q)$) and therefore, to lowest order in $\omega$,
\begin{align}
\mathrm{Im} \left[ - \chi^R_F(q,\omega) \right]  &\simeq  A\hbar \omega \int^{+W}_{-W} d\varepsilon
\int dk \, \rho(\varepsilon) \notag\\
&\times \rho'\left( \varepsilon - \frac{\hbar^2}{2m^*_F}(q^2 + 2k q) \right) \notag\\
& \times \theta\left(\epsilon_F - \varepsilon - \frac{\hbar^2 k^2}{2m^*_F}\right).
\label{eq:constrain}
\end{align}
In order to perform the integration over $k$, we define from the constraints imposed by the Heaviside step function in Eq.~\eqref{eq:constrain},   $k_F(\varepsilon) = \sqrt{\frac{2m^*_F}{\hbar^2}(\epsilon_F - \varepsilon)}$ for
$\varepsilon < \epsilon_F$, and note that
\begin{multline}
\int_{-k_{F}(\varepsilon)}^{+k_{F}(\varepsilon)} dk\,\partial_{\varepsilon}\rho\Big(\varepsilon-\frac{\hbar^2}{2m^*_{F}}(q^2+2kq)\Big)\\
=-\frac{m^*_{F}}{\hbar^2q}\int_{-k_{F}(\varepsilon)}^{+k_{F}(\varepsilon)} dk\,\partial_{k}\rho\Big(\varepsilon-\frac{\hbar^2}{2m^*_{F}}(q^2+2kq)\Big)\nonumber\\
\\=-\frac{m^*_{F}}{\hbar^2q}\Bigg[\rho\left(\varepsilon-\frac{\hbar^2}{2m^*_{F}}(q^2+2k_{F}(\varepsilon)q)\right)\nonumber\\
-\rho\Big(\varepsilon-\frac{\hbar^2}{2m^*_{F}}(q^2-2k_{F}(\varepsilon)q)\Big)\Bigg]
\end{multline}
Thus, the expression is  simplified and only the integration over $\varepsilon$  remains:
\begin{multline}
\mathrm{Im}[-\chi^{R}(q,\omega)]\simeq A\hbar\omega \Big(-\frac{m^*_{F}}{\hbar^2q}\Big)
\int_{-W}^{+W}d\varepsilon\, \theta(\epsilon_F - \varepsilon) \rho(\varepsilon)\\
\times  \Bigg[\rho\Big(\varepsilon-\frac{\hbar^2}{2m^*_{F}}(q^2+2k_{F}(\varepsilon)q)\Big)
-\rho\Big(\varepsilon-\frac{\hbar^2}{2m^*_{F}}(q^2-2k_{F}(\varepsilon)q)\Big)\Bigg] \label{eq:imchif}
\end{multline}
This expression can be numerically evaluated (cf. Fig.~\ref{fig:imchi}). However, for $q \to 0$, further analytical progress is possible   by noting that
$\rho(\varepsilon) \rho^{\prime}(\varepsilon) = \frac{1}{2}d [\rho(\varepsilon)]^2/d\varepsilon$, and hence,
\begin{align}
\mathrm{Im} \left[ - \chi^R_F(q \to 0,\omega) \right] &\simeq  A \hbar \omega    \int^{\min\{+W,\epsilon_F\}}_{-W} d\varepsilon
\, k_F(\varepsilon)
 \frac{d [\rho(\varepsilon)]^2}{d\varepsilon}.
\end{align}
From which, upon integration by parts, we obtain:

\begin{multline}
\mathrm{Im} \left[ - \chi_F^R(q \to 0,\omega) \right]  \simeq  A \hbar \omega \Big\{
-  [\rho(-W)]^2 k_F(-W) \\
  +[\rho(\min\{W,\epsilon_F\})]^2  k_F(\min\{W,\epsilon_F\}) \\
  + \frac{m^*_{F}}{\hbar^2} \int^{\min\{W,\epsilon_F\}}_{-W}
 d\varepsilon\, \frac{\left[\rho\left(\varepsilon\right)\right]^2  }{k_F(\varepsilon)} \Big\} \\
= A \hbar \omega   \frac{m^*_{F}}{\hbar^2} \int\limits^{\min\{+W,\epsilon_F\}}_{-W} d\varepsilon\, \frac{\left[\rho\left(\varepsilon\right)\right]^2  }{k_F(\varepsilon)}, \label{eq:back}
\end{multline}
Hence, by direct numerical evaluation of the above expression we see that it is  not singular, which
 implies that the $q\sim 0$ term (denoted $S^f_D$ in Eq.~\eqref{eq:sdissf}), can be neglected.  In general, using Eq.~\ref{eq:imchif} to evaluate $\mathrm{Im} \left[ -\chi^{R}_F(q,\omega)\right]$ for finite $q$, we find it is also a nonsingular function of $q$  in
 the neighborhood of  $q = 2 k^B_F = 2\pi \rho^0$. The results of a numerical evaluation of the integrals in equations~\eqref{eq:imchif} 
 and \eqref{eq:back} are displayed in Fig.~\ref{fig:imchi}.

Finally, the real part of the response function is given by:
\begin{align}
&\mathrm{Re}[\chi_F^R(q,\omega)]=A\int_{-\infty}^{\infty}d\varepsilon d\varepsilon' \rho(\varepsilon)\rho(\varepsilon')\nonumber\\
&\int\frac{dk}{2\pi}n_{k,\varepsilon}P\Bigg[\frac{1}{\hbar\omega+\varepsilon-\varepsilon' -\epsilon_{\parallel}(k+q)+\epsilon_{\parallel}(k)}+(\omega\to-\omega)\Bigg]\nonumber\\
&=A\int_{-W}^{\mathrm{min}[W,\epsilon_{F}]}d\varepsilon \rho(\varepsilon)\int \frac{dk}{2\pi}\,\theta\Big(\epsilon_{F}-\varepsilon-\frac{\hbar^2k^2}{2m^*_F}\Big)\nonumber\\
&\times\int_{-W}^{\mathrm{min}[W,\epsilon_{F}]}d\varepsilon' P\Bigg[\frac{\rho(\varepsilon')}{\hbar\omega+\varepsilon-\varepsilon' -\epsilon_{\parallel}(k+q)+\epsilon_{\parallel}(k)}\nonumber\\
&\qquad\qquad\qquad\qquad\qquad\qquad\qquad\qquad+(\omega\to-\omega)\Bigg]\nonumber\\
&=A\int_{-W}^{\mathrm{min}[W,\epsilon_{F}]}d\varepsilon \rho(\varepsilon) \int_{-k_{F}(\varepsilon)}^{k_{F}(\varepsilon)}  \frac{dk}{2\pi}\,\nonumber\\
&\times\int_{-W}^{\mathrm{min}[W,\epsilon_{F}]}d\varepsilon' P\Bigg[\frac{\rho(\varepsilon')}{E-\varepsilon' }+(\omega\to-\omega)\Bigg]
\label{prev}
\end{align}
where we have introduced $E=\hbar\omega+\varepsilon-\epsilon_{\parallel}(k+q)+\epsilon_{\parallel}(k)$.
Furthermore, by using the well-known Kramers-Kronig relations that connect the real and imaginary part of any complex function which is analytic in the upper half plane:
\begin{align}
\mathrm{Re}[G^{R}(\mathbf{R} =0,\varepsilon)]=-P\int_{-\infty}^{\infty}\frac{d\varepsilon'}{\pi}\frac{\mathrm{Im}[G^{R}(\mathbf{R}=\boldsymbol{0},\varepsilon')]}{\varepsilon-\varepsilon'}
\end{align}
we have that:
\begin{align}
P\int d\varepsilon' \frac{\rho(\varepsilon')}{E-\varepsilon' }&=P\int d\varepsilon' \frac{(-1/\pi)\mathrm{Im}[G^{R}(\mathbf{R}=\boldsymbol{0},\varepsilon')]}{E-\varepsilon' }\nonumber\\
&=\mathrm{Re}[G^{R}(\mathbf{R}=\boldsymbol{0},E)]
\end{align}
Then, we can rewrite equation (\ref{prev}), so that:
\begin{align}
&\mathrm{Re}[\chi_F^R(q,\omega)]=A\int_{-W}^{\mathrm{min}[W,\epsilon_{F}]}d\varepsilon \rho(\varepsilon)\nonumber\\
& \times\int_{-k_{F}(\varepsilon)}^{k_{F}(\varepsilon)}  \frac{dk}{2\pi}\,\Bigg[ \mathrm{Re}[G^{R}(\mathbf{R}=\boldsymbol{0},E)] +(\omega \to-\omega)\Bigg]\nonumber\\
&= \frac{A}{2\pi} \int\limits^{\min\{W,\epsilon_F\}}_{-W} d\varepsilon\, \rho(\varepsilon)\notag\\
&\times \int^{+k_F(\varepsilon)}_{-k_F(\varepsilon)} dk \, \Bigg[ g\left( \hbar \omega +\varepsilon  - \frac{\hbar^2 q^2}{2m^*_F} - \frac{\hbar^2 k q}{m^*_F} \right)\nonumber\\
&\qquad\qquad\qquad+ g\left( - \hbar \omega + \varepsilon  - \frac{\hbar^2 q^2}{2m^*_F} - \frac{\hbar^2 k q}{m^*_F}\right) \Bigg]
\end{align}
where $g(\varepsilon)=-\mathrm{Re}[G^{R}(\mathbf{R}=\boldsymbol{0},\varepsilon)]$ is the Hilbert transform of the density of states in a 2D square lattice modeled by a tight-binding approximation~\cite{Economou}:
\begin{align}
g(\varepsilon) = P \int d\varepsilon^{\prime} \: \frac{\rho(\varepsilon^{\prime})}{\varepsilon^{\prime}-\varepsilon} = \left\{
\begin{array}{lr}
-\frac{2}{\pi \varepsilon} \mathcal{K}\left(\frac{\varepsilon}{W} \right) & \mbox{for} |\varepsilon| \ge W, \\
-\frac{\mathrm{sgn}(\varepsilon)}{\pi W} \mathcal{K}\left(\frac{\varepsilon}{W} \right) & \mbox{for} |\varepsilon| <  W,
\end{array}
\right.
\end{align}
In particular, the \emph{static} limit $\omega = 0$ reads:
\begin{align}
&\chi_s(q) = \mathrm{Re}\left[ \chi_F^R(q,\omega = 0) \right] = \frac{A}{\pi} \int\limits^{\min\{W,\epsilon_F\}}_{-W} d\varepsilon\, \rho(\varepsilon)\nonumber\\
&\times\int^{+k_F(\varepsilon)}_{-k_F(\varepsilon)} dk \, g\left(\varepsilon  - \frac{\hbar^2 q^2}{2m^*_F} - \frac{\hbar^2 k q}{m^*_F} \right)
\end{align}
Therefore, it is possible to perform the calculation of the previous expression. \\

%
At low frequencies  ($\hbar\omega \ll \mu_B < \epsilon_F$) we shall approximate the response function of the Fermi gas
by the two first terms in the series about $\omega = 0$, i.e.
\begin{equation}
\chi_F^R(q,\omega) \simeq \chi_s(q) -  i \pi  \omega D(q),
\label{eq:gammaR}
\end{equation}
where $S(q) = \mathrm{Im} \left[- \chi_F^R(q)\right]/(\omega\pi)$
Finally, we make use of the  spectral properties of relating the retarded response function to its analytical continuation
to imaginary frequencies derived in the  Appendix~\ref{app:relating}:
\begin{equation}
\chi(q,\omega_n) =  - \int \frac{d\omega}{\pi} \frac{\mathrm{Im} \: \chi(q,\omega)}{i\omega_n - \omega}.
\end{equation}
In particular, the static limit $\omega_n = 0$ corresponds to:
\begin{equation}
\chi_s(q) = \chi(q,0) =   \int \frac{d\omega}{\pi} \frac{\mathrm{Im} \: \chi(q,\omega)}{\omega}.
\end{equation}
Adding and subtracting the static part,
\begin{align}
\chi(q,\omega_n) &=  \chi_s(q) - \int \frac{d\omega}{\pi} \,\mathrm{Im} \: \chi(q,\omega)
\left[\frac{1}{i\omega_n - \omega} + \frac{1}{\omega}\right]  \nonumber\\
&=  \chi_s(q) - \int \frac{d\omega}{\pi} \, \frac{\mathrm{Im} \: \chi(q,\omega)}{\omega} \left[ \frac{i\omega_n}{i\omega_n - \omega}\right].
\end{align}
and recalling that
\begin{align}
&\chi(q,\tau) = \int \frac{d\omega_n}{2\pi} e^{-i\omega_n \tau} \, \chi(q,\omega_n) \nonumber\\
&=  \chi_s(q) \delta(\tau) -
\int \frac{d\omega}{2\pi} \frac{\mathrm{Im} \: \chi(q,\omega)}{\omega}  \int^{+\infty}_{-\infty} \frac{d\omega_n}{\pi} \left[ \frac{i\omega_n \:
e^{-i\omega_n\tau}}{i\omega_n - \omega}
\right]  \nonumber\\
& =  \chi_s(q) \delta(\tau) -
\int^{+\infty}_0 \frac{d\omega}{\pi} \frac{\mathrm{Im} \: \chi(q,\omega)}{\omega} \nonumber\\
&\qquad\qquad\qquad \times\int^{+\infty}_{-\infty}  \frac{d\omega_n}{2\pi} \left[ \frac{i\omega_n}{i\omega_n - \omega} +  \frac{i\omega_n}{i\omega_n + \omega}
\right] \: e^{-i\omega_n\tau}.
\end{align}
Thus, upon performing the above integral over $\omega_n$ using Cauchy's theorem, the following expression is obtained:
\begin{align}
\chi(q,\tau) = \chi_s(q) \delta(\tau) + \int^{+\infty}_0 \frac{d\omega}{\pi} \, e^{-\omega |\tau|} \: \mathrm{Im} \: \chi_F^R(q,\omega).
\end{align}
Hence, introducing Eq.~(\ref{eq:gammaR}) in the expression above,
\begin{equation}
\chi(q,\tau) \simeq \chi_s(q) \: \delta(\tau) - \frac{D(q)}{(|\tau| + \tau_c)^2}
\end{equation}
The first term describes the short time behavior, which is dominated by  \textit{screening}, whereas the second term describes the long time behavior, which is dominated by \textit{dissipation}.


\section{RG analysis at half-filling} \label{app:rg}

In order to obtain the RG flow equations, we consider the functional integral representation of the
partition function:
\begin{equation}
Z(\tau_c) = \int \left[ d \phi\right] \, e^{-S[\phi]}, \label{eq:partition}
\end{equation}
where
\begin{equation}
S[\phi] = S_0[\phi] + S_{\mathrm{int}}[\phi],
\end{equation}
$S_0[\phi]$ being the Gaussian part of the action (the first term in Eq.~\ref{eq:sb0}). When
writting \eqref{eq:partition}, we have made explicit the dependence of the partition function on the short-distance
cut-off $a_0\simeq v \tau_c$.  Note, however, that (up to a multiplicative constant), the partition function is independent
of the cut-off, and we will base our subsequent analysis on this fact.  For a general perturbation $S_{\text{int}}[\phi]$
we cannot compute the partition function exactly. Thus, we resort to a perturbative  expansion of $Z[(1+\delta \ell) \tau_c]$
(where $\delta \ell \ll 1$) in powers  of $S_{\text{int}}$:
\begin{align}
Z[(1+\delta \ell)\tau_c ] &= Z_0[(1+\delta \ell)\tau_c] \Big\{ 1 - \langle S_{\text{int}}[\phi]\rangle  \\
&  + \frac{1}{2} \langle S^2_{\text{int}}[\phi]\rangle + \cdots \Big\}
\end{align}
To deal with this expansion it is convenient to define the normal ordered vertex operators:
\begin{equation}
:e^{2 p i\phi(\mathbf{x})}: \, = \frac{1}{a_0^{p^2 K}} e^{2 p i\phi(\mathbf{x})}
\end{equation}
where  $\mathbf{x} = (v\tau,x)$ the limit $a_0  \to 0$ is implicitly understood. Then, when inserted in an expectation value,
we have the following operator product expansions (OPE):
\begin{align}
&:e^{2ip\phi(\mathbf{r})}::e^{-2ip\phi(\mathbf{r}')}:\nonumber\\
&=\frac{1}{|\mathbf{r}-\mathbf{r}'|^{2p^2K}}:[1+2ip(\mathbf{r}-\mathbf{r}')\nabla\phi(\mathbf{R})\nonumber\\
&\qquad\qquad\qquad\qquad-2p^2[(\mathbf{r}-\mathbf{r}')\nabla\phi(\mathbf{R})]^2+\cdots]:\\
&:e^{2ip\phi(\mathbf{r})}::e^{2ip\phi(\mathbf{r}')}:=a_{0}^{2p^2K}:e^{4ip\phi(\mathbf{R})}:+\cdots
\end{align}
where $\mathbf{r}=(v\tau,x)$, $\mathbf{R}=(\mathbf{r}-\mathbf{r}')/2$ $\nabla=((1/v)\partial\tau, \partial x)$ and $a_{0}=v\tau_c$ is a short-distance cut-off. Next, let us consider the partition function at the scale $(1+\delta l)a_{0}$, where $\delta l >0$ and $\delta l \ll 1$:
\begin{align}
Z[(1+\delta l)a_{0}]=Z_{0}[(1+\delta l)a_{0}]\Big\{1-\langle S_{int}\rangle + \frac{1}{2!}\langle S_{int}^2 \rangle +\cdots\Big\}
\end{align}
where
\begin{align}
&S_{u}[\phi]=-\frac{g_{u}}{\pi a_{0}^{2-4K}}\int dx d\tau :\cos{4\phi(\mathbf{r})}:,\\
&S_{D}[\phi]=-\frac{g_{D}}{a^{1-2K}_{0}}\int_{|\mathbf{r}-\mathbf{r}'|>a_{0}} d\mathbf{r} d\mathbf{r}' \frac{\delta(x-x')}{|\mathbf{r}-\mathbf{r}'|^2}:\cos{2\phi(\mathbf{r}}):\notag\\
&\quad \quad \times :\cos{2\phi(\mathbf{r}')}:
\end{align}
where we have normal ordered the vertex operators. 

\subsection{First order terms}
Now, let us consider the first order term $\langle S_{int}\rangle=\langle S_{u} \rangle + \langle S_{D} \rangle$:
\begin{align}
-\langle S_{u} \rangle=+\frac{g_{u}(l + \delta l)}{\pi [(1+\delta l) a_{0}]^{2-4K}}\int d\mathbf{r} \langle:\cos{4\phi(\mathbf{r})}:\rangle
\end{align}
When compared with the same operator at the scale $a_{0}$, we find that:
\begin{align}
&\qquad\frac{g_{u}(l+ \delta l)}{ [(1+\delta l) ]^{2-4K}}=g_{u}(l) \Longrightarrow \nonumber\\
&g_{u}(l+ \delta l)=g_{u}(l) [1-(2-4K)\delta l],
\end{align}
which inmediately leads to the differential equation:
\begin{align}
\frac{dg_{u}(l)}{dl}=(2-4K)g_{u}(l)
\label{eq:gudiff}
\end{align}
Next, we consider:
\begin{align}
&-\langle S_{D} \rangle=+\frac{g_{D}(l+ \delta l)}{\pi [(1+\delta l) a_{0}]^{1-2K}}\times\nonumber\\
&\int_{|\mathbf{r}-\mathbf{r}'|>a_{0}(1+\delta l)} d\mathbf{r} d\mathbf{r}'  \frac{\delta(x-x')}{|\mathbf{r}-\mathbf{r}'|^2}\langle:\cos{2\phi(\mathbf{r}})::\cos{2\phi(\mathbf{r}')}:\rangle
\end{align}
To bring this expression to a form which can be compared with the same expression at the cut-off scale $a_{0}$, we first split the integral on $\mathbf{r}$ and $\mathbf{r}'$ as follows:
\begin{align}
&\int_{|\mathbf{r}-\mathbf{r}'|>a_{0}(1+\delta l)} d\mathbf{r} d\mathbf{r}' \cdots\nonumber\\
&=\int_{|\mathbf{r}-\mathbf{r}'|>a_{0}}d\mathbf{r} d\mathbf{r}' \cdots -\int_{a_{0}(1+\delta l) > |\mathbf{r}-\mathbf{r}'|>a_{0}}d\mathbf{r} d\mathbf{r}' \cdots \label{eq:split}
\end{align}
Thus, from the first term int the right hand-side of the above equation, we have that:
\begin{align}
&+\frac{g_{D}(l+ \delta l)}{\pi [(1+\delta l) a_{0}]^{1-2K}}\times\nonumber\\
&\int_{|\mathbf{r}-\mathbf{r}'|>a_{0}} d\mathbf{r} d\mathbf{r}'  \frac{\delta(x-x')}{|\mathbf{r}-\mathbf{r}'|^2}\langle:\cos{2\phi(\mathbf{r}})::\cos{2\phi(\mathbf{r}')}:\rangle
\end{align}
Hence, following the same procedure as before:
\begin{align}
&\frac{g_{D}(l+ \delta l)}{ [(1+\delta l) ]^{1-2K}}=g_{D}(l) \Longrightarrow \nonumber\\
&\quad\frac{dg_{D}(l)}{dl}=(1-2K)g_{D}(l),
\end{align}
Next, we take up the contribution from the second term in Eq.~\ref{eq:split}:
\begin{align}
&-\frac{g_{D}(l+ \delta l)}{ a_{0}^{1-2K}}\int_{a_{0}(1+\delta l) > |\mathbf{r}-\mathbf{r}'|>a_{0}}d\mathbf{r} d\mathbf{r}'  \frac{\delta(x-x')}{|\mathbf{r}-\mathbf{r}'|^2}\times\nonumber\\
&\qquad\qquad\qquad\qquad\qquad\langle:\cos{2\phi(\mathbf{r}})::\cos{2\phi(\mathbf{r}')}:\rangle\nonumber\\
&=-\frac{g_{D}(l+ \delta l)}{2a_{0}^{1-2K}}\int_{a_{0}(1+\delta l) > |\mathbf{r}-\mathbf{r}'|>a_{0}}d\mathbf{r} d\mathbf{r}'  \frac{\delta(x-x')}{|\mathbf{r}-\mathbf{r}'|^{2+2K}}\times\nonumber\\
&\qquad\qquad\quad\qquad\langle:\Big[1-2[(\mathbf{r}-\mathbf{r}')\nabla\phi(\mathbf{R})]^2+\cdots\Big]:\rangle\nonumber\\
&=-\frac{g_{D}(l+ \delta l)}{2a_{0}^{1-2K}}\int_{a_{0}(1+\delta l) > |\mathbf{r}-\mathbf{r}'|>a_{0}}d\mathbf{r} d\mathbf{r}'  \frac{\delta(x-x')}{|\mathbf{r}-\mathbf{r}'|^2}\times\nonumber\\
&\qquad\qquad\quad\qquad  a_{0}^{2k}\langle:\cos{4\phi(\mathbf{R})}:\rangle
\label{eq:recall}
\end{align}
Introducing $\mathbf{u}=\mathbf{r}-\mathbf{r}'$ leads to
\begin{align}
&-\frac{g_{D}(l)\delta l}{ a_{0}^{1-4K}}\int_{a_{0}(1+\delta l) > |\mathbf{u}|>a_{0}}d\mathbf{u}\frac{\delta(u_x)}{|\mathbf{u}|^2}\nonumber\\
&=\Big[\int_{a_{0}}^{a_{0}(1+\delta l)}\frac{du_x}{u_x^2}\Big]\frac{1}{a_{0}^{1-4K}}=\frac{\delta l}{a_{0}^{2-4K}}
\end{align}
Hence, the second term in Eq.~(\ref{eq:recall}) yields:
\begin{align}
-\frac{g_{D}(l)\delta l}{a_{0}^{2-4K}}\int d\mathbf{r} :\cos{4\phi(\mathbf{r})}:
\end{align}
Therefore, the flow equation for $g_{u}(l)$ (i.e. Eq.~\ref{eq:gudiff})  must be modified to:
\begin{align}
&\frac{g_{u}(l+ \delta l)}{ [(1+\delta l) ]^{2-4K}}-\pi g_{D}(l)\delta l=g_{u}(l) \nonumber\\
&g_{u}(l+ \delta l)=[1+(2-4K)\delta l]g_{u}(l)+\pi g_{D}(l)\delta l\nonumber\\
&\quad\frac{dg_{u}(l)}{dl}=(2-4K)g_{u}(l)-\pi g_D(l)
\end{align}
Finally, it is necessary to consider the first term in Eq.~(\ref{eq:recall}). To this end, we need to consider the following integral with $\mathbf{r}-\mathbf{r}'=\mathbf{u}=(u\tau,ux)$:
\begin{align}
&-\frac{g_{D}(l)\delta l}{ a_{0}^{1-2K}}\int_{a_{0}(1+\delta l) > |\mathbf{u}|>a_{0}}d\mathbf{u}\frac{\delta(ux)}{|\mathbf{u}|^{2+2K}}\nonumber\\
&=\Big[\int_{a_{0}}^{a_{0}(1+\delta l)}\frac{du}{u^{2K}}\Big]\frac{1}{a_{0}^{1-4K}}\nonumber\\
&=\Big[\int_{a_{0}}^{a_{0}(1+\delta l)}d\Big(\frac{u}{a_{0}}\Big)\Big(\frac{a_{0}}{u}\Big)^{2K}\Big]=\delta l
\end{align}
Thus, a term of the following form is generated:
\begin{align}
\frac{2g_{D}(l)}{v(l)}\delta l \int dxd\tau \langle (\partial_{\tau} \phi)^2\rangle
\end{align}
Note that this term has the same form of the $\frac{1}{2\pi Kv}(\partial_\tau \phi)^2$ operator in the Gaussian action, $S_{0}[\phi]$. When re-exponentiated, we fined that, upon comparing with the same action at the cut-off scale as:
\begin{align}
\frac{1}{2\pi K(l + \delta l) v (l + \delta l) }-2\frac{g_{D}(l)}{v(l)}\delta l = \frac{1}{2\pi K(l) v(l)}
\end{align}
Hence,
\begin{align}
&\frac{1}{K(l + \delta l) v (l + \delta l) }= \frac{1}{2\pi K(l) v(l)} + 4\pi \frac{g_{D}(l)}{v(l)}\delta l \Longrightarrow \nonumber\\
&\qquad\qquad\qquad \frac{d}{dl}\Big(\frac{1}{Kv}\Big)=\frac{4\pi g_{D}}{v}
\end{align}
Furthermore, the coefficient of $\int dxd\tau (\partial_{x}\phi)^2$ is not renormalized:
\begin{align}
&\frac{v(l+\delta l)}{K(l+\delta l)}=\frac{v(l)}{K(l)} \Longrightarrow \frac{d}{dl}\Big(\frac{v}{K}\Big)=0
\end{align}
From these equations we can extract the RG flow equations for $k$ and $v$:
\begin{align}
&\frac{1}{K}\frac{dv}{dl}+ v\frac{d}{dl}\Big(\frac{1}{K}\Big)=0 \label{eq:AA}\\
&\frac{1}{Kv^2}\frac{dv}{dl}+ \frac{1}{v}\frac{d}{dl}\Big(\frac{1}{K}\Big)=\frac{4\pi g_{D}}{v} \Longrightarrow \nonumber\\
&\frac{1}{K}\frac{dv}{dl}+ v\frac{d}{dl}\Big(\frac{1}{K}\Big)=4\pi g_{D}v \label{eq:BB}
\end{align}
Thus, adding Eq.~(\ref{eq:AA}) and Eq.~(\ref{eq:BB}):
\begin{align}
2\frac{d}{dl}\Big(\frac{1}{K}\Big)=4\pi g_{D}\Longrightarrow \frac{d}{dl}\Big(\frac{1}{K}\Big)=2\pi g_{D}.
\end{align}
\subsection{Second order terms}
After considering the first order contributions, we need to take up the second order:
\begin{align}
\frac{1}{2!}\langle S_{int}^2 \rangle=\frac{1}{2!}\langle S_{u}^2 [\phi] \rangle +\cdots
\end{align}
We do not consider terms of order $g_{u}g_{D}$ or $g_{D}^2$ because $g_{D} \propto g_{BF}^2$ is already second order and $g_{u} \ll 1$ is considered small. Thus, taking:
\begin{align}
&\frac{1}{2!}\langle S_{u}^2 [\phi] \rangle=\frac{1}{2}\Bigg(\frac{g_{u}(l+ \delta l)}{\pi [(1+\delta l) a_{0}]^{2-4K}}\Bigg)^2\times\nonumber\\
&\int_{|\mathbf{r}-\mathbf{r}'|>a_{0}(1+\delta l)} d\mathbf{r}d\mathbf{r}' \langle:\cos{4\phi(\mathbf{r})}::\cos{4\phi(\mathbf{r}')}:\rangle
\end{align}
Again, we split the integral as in Eq.~\eqref{eq:split}, which leads to:
\begin{align}
&-\frac{1}{2}\Bigg(\frac{g_{u}(l)}{\pi a_{0}^{2-4K}}\Bigg)^2\times\nonumber\\
&\int_{a_{0}(1+\delta l)>|\mathbf{r}-\mathbf{r}'|>a_{0}} \frac{1}{2}d\mathbf{r}d\mathbf{r}' \langle:\Big(1-8[(\mathbf{r}-\mathbf{r}')\nabla\phi(\mathbf{R})]^2+\cdots\Big):\rangle\nonumber\\
&\qquad\qquad\qquad\qquad\qquad\qquad\qquad\qquad\qquad\times\frac{1}{|\mathbf{r}-\mathbf{r}'|^{8K-4}}\nonumber\\
&= \frac{4}{\pi^2}\int d\mathbf{R} \langle |\nabla_{\mathbf{R}}\phi(\mathbf{R})|^2 \rangle \nonumber\\
&\int_{a_{0}(1+\delta l)>|\mathbf{r}-\mathbf{r}'|>a_{0}} \Big(\frac{a_{0}}{|\mathbf{u}|}\Big)^{8K-4}\cos^2\phi+\cdots\nonumber\\
&=\frac{2}{\pi}g_{u}^2\delta l\int d\mathbf{r} \langle :\Big(\nabla\phi(\mathbf{r})\Big)^2 : \rangle + \cdots
\end{align}
Thus, we need to revise our previously derived equations for the renormalization of the Gaussian action parameters:
\begin{align}
&\frac{1}{2\pi K(l + \delta l) v (l + \delta l) }-2\frac{g_{D}(l)}{v(l)}\delta l -2\frac{g_{u}^2(l)}{\pi v(l)}\delta l = \frac{1}{2\pi K(l) v(l)}\\
&\frac{v(l+\delta l)}{2\pi K(l+\delta l)}-\frac{2v(l)g_{u}^2(l)\delta l}{\pi}=\frac{v(l)}{2\pi K(l)}
\end{align}
Hence,
\begin{align}
&\frac{d}{dl}\Big(\frac{1}{Kv}\Big)=\frac{4\pi g_{D}}{v} + 4 \pi \frac{g_{u}^2}{v}\\
&\frac{d}{dl}\Big(\frac{v}{K}\Big)=4v g_{u}^2
\end{align}
\begin{align}
&\frac{1}{v}\frac{d}{dl}\Big(\frac{1}{K}\Big)-\frac{1}{Kv^2}\frac{dv}{dl}=\frac{4\pi}{v}[g_{D}+\frac{g_{u}^2}{\pi} ]\\
&v\frac{d}{dl}\Big(\frac{1}{K}\Big)+\frac{1}{K}\frac{dv}{dl}=4vg_{u}^2
\end{align}
Hence, the RG flow equations for both K and v:
\begin{align}
&\frac{d}{dl}\Big(\frac{1}{K}\Big)-\frac{1}{Kv^2}\frac{dv}{dl}=\frac{4\pi}{v}[g_{D}+\frac{g_{u}^2}{\pi} ]\\
&\frac{d}{dl}v=-2\pi g_{D}Kv
\end{align}

Finally, in the analysis of the second order contributions, we need to consider the term:
\begin{align}
&O(g_{u}g_{D})=\frac{2g_{u}(l+\delta l)}{\pi [a_{0}(1+\delta l)]^{4-2K}}\frac{g_{D}(l+\delta l)}{[a_{0}(1+\delta l)]^{1-2K}}\nonumber\\
&\times \int_{*} d \mathbf{r}_{1} d \mathbf{r}_{2} d \mathbf{r}_{3}\frac{\delta(x_{1}-x_{2})}{|\mathbf{r}_{1}-\mathbf{r}_{2}|^2}\langle:\cos{4\phi(\mathbf{r}_{1})}\cos{4\phi(\mathbf{r}_{2})}\cos{4\phi(\mathbf{r}_{3})}:\rangle
\end{align}
where the star ($*$) under the integral means that: $|\mathbf{r}_{1}-\mathbf{r}_{2}|>a_{0}(1+\delta l)$, $|\mathbf{r}_{1}-\mathbf{r}_{3}|>a_{0}(1+\delta l)$, and $|\mathbf{r}_{2}-\mathbf{r}_{3}|>a_{0}(1+\delta l)$.
Let us consider the contribution resulting from the OPE when $\mathbf{r}_{1}\to\mathbf{r}_{2}$ (or equivalently $\mathbf{r}_{1}\to\mathbf{r}_{3}$):
\begin{align}
:\cos{4\phi(\mathbf{r}_{1})}::\cos{4\phi(\mathbf{r}_{2})}:=\frac{1}{2|\mathbf{r}-\mathbf{r}'|^{4K}}:\cos{2\phi(\mathbf{R})}: +\cdots
\end{align}
Hence, as the above factor of 2 is cancelled by the two possible contractions $\mathbf{r}_{1}\to\mathbf{r}_{2}$ and $\mathbf{r}_{2}\to\mathbf{r}_{3}$:
\begin{align}
&O(g_{u}g_{D})=-\frac{2g_{u}(l+\delta l)}{\pi [a_{0}(1+\delta l)]^{4-2K}}\frac{g_{D}(l+\delta l)}{[a_{0}(1+\delta l)]^{1-2K}}\nonumber\\
&\times \int_{a_{0}(1+\delta l)>|\mathbf{\rho}|>a_{0} }d \mathbf{\rho} \frac{1}{|\mathbf{\rho}|^{4K}} \int d \mathbf{R} d \mathbf{r}_{3}\frac{\delta(X-x_{3})}{|\mathbf{R}-\mathbf{r}_{3}|^2}\nonumber\\
&\times\langle:\cos{4\phi(\mathbf{R})}\cos{4\phi(\mathbf{r}_{3})}:\rangle+\cdots\nonumber\\
&=-\frac{4g_{u}(l)g_{D}(l)}{\pi [a_{0}(1+\delta l)]^{1-2K}}\delta l\nonumber\\
&\int d\mathbf{r} d\mathbf{r}'  \frac{\delta(x-x')}{|\mathbf{r}-\mathbf{r}'|^2}\langle:\cos{2\phi(\mathbf{r}})::\cos{2\phi(\mathbf{r}')}:\rangle
\end{align}
Therefore, we obtain the following differential equation:
\begin{align}
&\frac{g_{D}(l+\delta l)}{[a_{0}(1+\delta l)]^{1-2K}}-\frac{4g_{u}(l)g_{D}(l)\delta l}{ [a_{0}(1+\delta l)]^{1-2K}}=\frac{g_{D}(l)}{a_{0}^{1-2K}} \Longrightarrow \nonumber\\
&\qquad\qquad\frac{g_{D}}{dl}=(1-2K)g_{D} + 4g_{D}g_{u}
\end{align}
Thus, the complete set of RG flow equations reads:
\begin{align}
&\frac{dv}{dl}=-4\pi g_{D}Kv \label{eq:RG0} \\
&\frac{dK}{dl}=-(4 g_{u}^2 + 2\pi g_{D})K^2 \label{eq:RG1} \\
&\frac{dg_{u}}{dl}=(2 -4K)g_{u}+\pi g_{D} \label{eq:RG2} \\
&\frac{dg_{D}}{dl}=(1 -2K)g_{D}+4 g_{D}g_{u} \label{eq:RG3}
\end{align}

\section{SCHA} \label{app:scha}

We have adopted a variational self-consistent harmonic approximation (SCHA) by choosing a trial effective action such as in Eq.~\eqref{eq:strial}.
To find the variational estimate of the free-energy we have to perform the averages of the effective action with respect to the trial effective action, by using $S_{0}$ (Eq.~(\ref{eq:sb0})), $S_{u}$ (Eq.~(\ref{eq:su})) with $p=2$ for half-lattice filling, $S_{D}^{b}$ (Eq.~(\ref{eq:sdissb})) and $S_{D}^{u}$ (Eq.~(\ref{eq:sdissu})). Thus, the variational free-energy $F_{var}$ that follows from Eq.~(\ref{eq:fvar}) will be:
\begin{align}
&F_{var}[G_v]=-\frac{T}{2}\int \frac{dq d\omega}{(2\pi)^2}\ln G_v(q,\omega) \nonumber\\
&+ T\Big[\frac{1}{2\pi K} (\frac{\omega^2}{v_s} + v_sq^2)\Big] G_v(q,\omega)\nonumber\\
&-T\frac{g_{u}}{a_0\tau_0}\int dx d\tau e^{-8\int\frac{dqd\omega}{(2\pi)^2}G_v(q,\omega)}\nonumber\\
&- T\frac{g_D}{a_0}\int dx \int_{|\tau-\tau'|>\tau_0} d\tau d\tau' \frac{e^{-4\int\frac{dqd\omega}{(2\pi)^2}[1-\cos \omega(\tau-\tau')]G_v(q,\omega)}}{|\tau-\tau'|^2}\nonumber\\
&- T\frac{g_D}{a_0}\int dx \int_{|\tau-\tau'|>\tau_0} d\tau d\tau' \frac{e^{-4\int\frac{dqd\omega}{(2\pi)^2}[1+\cos \omega(\tau-\tau')]G_v(q,\omega)}}{|\tau-\tau'|^2}\nonumber\\
&-\langle S_v \rangle_v
\label{eq:fvv}
\end{align}
Therefore,  requiring $\delta F_{var}[G_v]/\delta G_v=0$ yields:
\begin{align}
&\delta F_{var}[G_v]/\delta G_v=-\frac{T}{2}\int \frac{dq d\omega}{(2\pi)^2}\frac{1}{G_v(q,\omega)}\nonumber\\
&+ T\Big[\frac{1}{2\pi K} (\frac{\omega^2}{v_s} + v_sq^2)\Big] +T\frac{8g_{u}}{(2\pi)^2 a_0\tau_0}\alpha^2(\eta,\Delta,K)\nonumber\\
&+ T\frac{4g_D}{(2\pi)^2a_0}\Bigg[\alpha(\eta,\Delta,K)+\alpha^2(\eta,\Delta,K)\Bigg]=0
\label{eq:self}
\end{align}
where $\alpha(\eta,\Delta,K)=\Big[\frac{\eta K\pi + 2\sqrt{K\pi\Delta}}{4}\Big]^{2K}$.
By keeping the $\tau$-independent terms in the integrals in Eq.~(\ref{eq:fvv}) which yield the
leading contributions in $\omega$ to $G_{v}(q,\omega)$, leads to equations~(\ref{eq:scha1},\ref{eq:scha2}).

\end{document}